\documentclass[a4paper]{article}

\usepackage[english]{babel}
\usepackage{graphicx,amsmath,amsfonts,amssymb,slashed,upgreek,color,caption,amscd,cite,cancel}
\usepackage{framed}
\usepackage{multirow}
\usepackage{authblk}
\usepackage{graphicx,subfigure}
\usepackage{multicol}
\usepackage{a4wide}
\usepackage{fullpage}
\usepackage{epstopdf}
\usepackage{float} 
\usepackage[usenames,dvipsnames,svgnames,table]{xcolor}
\usepackage{bm} 
\usepackage{mathrsfs} 
\usepackage{relsize} 
\usepackage{multirow}
\usepackage{pdfpages} 
\usepackage{url}
\usepackage{bigints}

\definecolor{Myblue}{rgb}{0.,0.,0.8}
\usepackage{hyperref}
\hypersetup{colorlinks,citecolor=Myblue,linkcolor=black,filecolor=black,urlcolor=black}

\usepackage{geometry}
\def\m22{\mu_2^2}
\def\l33{\Lambda^3_3}

\def\m22{\mu_2^2}

\title{Flavour composition and entropy increase of cosmological neutrinos after decoherence}
\author[1,2]{Daniel Boriero}
\author[1]{Dominik J. Schwarz}
\author[3]{Hermano Velten}
\affil[1]{Fakult\"at f\"ur Physik, Universit\"at Bielefeld, Postfach 100131, 33501, Bielefeld, Germany}
\affil[2]{Centro de Ci\^encias Naturais e Humanas, Universidade Federal do ABC (UFABC), 09210-580, Santo Andr\'e, Brazil }
\affil[3]{Departamento de F\'isica, Universidade Federal de Ouro Preto (UFOP), 35400-000, Ouro Preto MG, Brazil}
\date{}

\begin{document}
\maketitle

\abstract{We propose that gravitational interactions of cosmic neutrinos with the statistically 
homogeneous and isotropic fluctuations of space-time leads to decoherence. 
This working hypothesis, which we describe by means of a Lindblad operator, is applied to the system of 
two- and three-flavour neutrinos undergoing vacuum oscillations and the consequences are 
investigated. As result of this decoherence we find that the neutrino entropy would increase as a function 
of initial spectral distortions, mixing angles and CP-violation phase. Subsequently we discuss the chances to discover 
such an increase observationally (in principle). We also present the expected flavour composition of the cosmic neutrino 
background after decoherence is completed. The physics of two- or three-flavour oscillation of cosmological neutrinos 
resembles in many aspects two- or three-level systems in atomic clocks, which were recently proposed by Weinberg for 
the study of decoherence phenomena.}

\tableofcontents

\section{Introduction}

The indirect evidence for the existence of a cosmic neutrino background (CNB) is a major achievement of 
observational cosmology \cite{Cyburt:2004yc,2011ApJS..192...18K,2016A&A...594A..13P}. Its direct detection 
remains an experimental challenge  \cite{PhysRev.128.1457,Betts:2013uya}. 
 
Active neutrinos are in thermal equilibrium in the early Universe. Their momentum spectra follow the Fermi-Dirac 
distribution, which is preserved after their decoupling --- even after they became non-relativistic. 
Non-instantaneous decoupling and neutrino heating by electron-positron annihilation 
introduce distortions to the perfect Fermi-Dirac spectrum for each neutrino flavour \cite{Dolgov2002363,Hannestad:2001iy}. 
These spectral distortions are smoothed out by neutrino oscillations, but do not disappear completely. 
These distortions in the neutrino spectra imply a small increase of the neutrino density, giving $N_{\rm eff} = 3.046$ 
in terms of effective number of neutrinos. In order to arrive at this result, one assumes that 
there are no primordial lepton-flavour asymmetries.

In this work we study the cosmic evolution and decoherence (i.e., the irreversible loss of quantum coherence) of 
active neutrino flavours from their decoupling (when the Universe 
had a temperature of $\sim 1$ MeV) to the present time. 
We ask the question what happens to the neutrino mass states when 
they become non-relativistic at late times. 
We propose in this work that the neutrino mass states undergo decoherence when the 
heaviest and second heaviest states become non-relativistic. 
The reasoning behind this hypothesis is that mass states in the non-relativistic regime have very different
group velocities and therefore travel on different paths in space-time. The isotropic and homogeneous universe 
described by the $\Lambda$CDM model is homogeneous only in a statistical sense. There are fluctuations of 
the gravitational potential that vary in position and time. Coherent states traveling through this space-time may 
accumulate enough differences in their paths to reach an irreversible trajectory, causing the loss of information 
and the respective decoherence. Corresponding to the decoherence, a time averaging of the phase of these 
states traveling through different paths in space-time would also result in a break up of time reversibility.
As a consequence of this hypothetical decoherence,
there is an associated increase of neutrino entropy. We also present the small corrections to the flavour distribution
of the CNB today, which were seeded by primordial spectral distortions. 
  
To track the evolution of the flavour content of individual neutrinos of given momentum we make use of the 
Wigner density matrix in a basis $\{\left| v_a \right\rangle\}$
\begin{equation}
\rho_p(t) = \sum_{a,b} \left| v_a \rangle \rho_{ab}(p,t) \langle v_b \right|,  
\end{equation}
where the entries $\rho_{ab} (t, p)$ 
are defined at time $t$ and momentum $p$.

To obtain the distribution function for an ensemble of neutrinos specified by the flavour state label $\alpha$,
we have to trace the projected density matrix $\rho_p$ and multiply by the appropriate momentum spectrum and spin degrees of freedom, 
\begin{equation}
f_\alpha (p,T) = f_\alpha^{\rm eq} (p,T) \mathrm{tr}\left[ \left| v_\alpha \rangle \langle v_\alpha \right| \rho_p \right],
\end{equation}
where $f_\alpha^{\rm eq}(p,T)$ is the equilibrium momentum spectrum at temperature $T$ in which we also account for 
the spin degrees of freedom.
The density matrix contains the flavour dependent spectral distortions, which are also momentum dependent.
Since we are only interested in the flavour composition of spectral distortions, we focus on the normalized 
density matrix spanned by the basis $\{\left| v_a \right\rangle\}$. Moreover, we drop the subindex 
$p$ that indicates the momentum dependence for simplicity, but all the equations are momentum dependent.
For the purpose of this work, the momentum dependence is trivial as the comoving momentum  is conserved 
after neutrino decoupling.

The time evolution of the density matrix $\rho$ is given by the von Neumann equation. In an 
expanding Friedman-Lema\^itre model, after neutrino decoupling and after electron-positron annihilation 
(i.e.\ for $T < m_e/3 \approx 0.2$ MeV), it reads\footnote{We use units in which 
$\hbar = c = k_\text{B} = 1$ and $M_{\rm P} = 1/\sqrt{8\pi G}$ is the reduced Planck mass.}
\begin{equation}
\imath D \rho = [{\mathcal H}, \rho], 
\label{vN}
\end{equation}
where ${\mathcal H}$ is the free Hamiltonian. The density matrix and the Hamiltonian are both functions 
of the cosmic scale factor $a$ and the modulus of the comoving neutrino momentum $q \equiv a p$. 
Both $\rho$ and $\mathcal H$ are hermitian and $\mathrm{tr} \rho = 1$. The differential operator 
\begin{equation}
D = \frac{d}{dt} = H(a) \frac{{\rm d}}{{\rm d}\ln a}, 
\end{equation} 
with $H$ denoting the Hubble expansion rate and with $a = 1$ today.
The cosmological redshift $z = 1/a - 1$.

A pure state is characterised by  $\mathrm{tr} \rho^2 = 1$, while for mixed states $\mathrm{tr} \rho^2 <1$.
Thus any physical system has $\mathrm{tr} \rho^2 \leq 1$. Since the von Neumann equation preserves 
$\mathrm{tr} \rho^2$, it cannot describe the process of decoherence. This follows as the von Neumann 
equation is a direct consequence of the unitary time evolution described by the Schr\"odinger equation.  

The density matrix is connected to the (von Neumann) entropy via
\begin{equation}
S = - \mathrm{tr}[\rho\ln\rho].  \label{eqentropy}
\end{equation}
The von Neumann entropy vanishes for pure states and is maximal for maximally mixed states. 
For a two (three) flavour state system, the maximally achievable family entropy (following Boltzmann) obviously 
is $S = \ln 2$ ($\ln 3$). Here we ignore the spin and momentum dependence of neutrinos. The von Neumann 
equation conserves this entropy and does not describe decoherence.

Here we assume that different mass states in the non-relativistic regime travel through 
different stochastic gravitational potentials, which leads effectively to a phase averaging. A time average 
along the neutrino path will consequently cause a definite decoherence. This loss of information cannot be recovered, 
it is not a de-phasing or kinetic decoherence that can be undone by a precise measurement. However, we will not 
model the different paths that the neutrino states travel and neither perform the time average, instead we will 
introduce a decoherence operator in the von Neumann equation encompassing the gravitational environmental 
fluctuations. The methodology of using an effective operator to introduce environmental sources of decoherence 
in an open quantum system is commonly adopted, including for neutrinos \cite{Oliveira:2014jsa,Gomes:2016ixi,Coelho:2017zes}.

The modified von Neumann equation with a proper operator to describe decoherence is given by
\begin{equation}
\imath D \rho^{\rm L} = [{\mathcal H}, \rho^{\rm L}] + \imath \sum_a \left[{\cal L}_a,\left[\rho^{\rm L},{\cal L}_a\right]\right] \ ,
\label{vNnew}
\end{equation}
which we refer to as Lindblad equation \cite{Lindblad1976}. The ${\cal L}_a$ are so-called 
Lindblad operators, arising from tracking or averaging environment dynamics. 
The decoherence term is responsible for the fact that the quantum system can develop dissipation and 
irreversibility and lose quantum coherence. As we will show shortly, the entropy for the density matrix $\rho^{\rm L}$, as well as for a time averaged density matrix over irreversible paths $\bar \rho$, is not constant anymore.

Many aspects of oscillating neutrinos in the early Universe have been discussed before. The focus of those 
works may be in the interplay with the primordial plasma \cite{Notzold:1987ik,PhysRevD.49.2710}, 
the matter effect \cite{Hannestad:2001iy,Volpe:2013jgr}, 
the role of a lepton asymmetry \cite{Wong:2002fa,Dolgov:2002ab,Barenboim:2016lxv}
or the back-reaction effects in primeval 
nucleosynthesis \cite{Mangano:2005cc,Saviano:2013ktj}. 
Important for this work are studies of distortions in 
the neutrino spectral distribution \cite{Kirilova:2002ss,Mangano:2005cc,Langacker1987589}. These works rely 
on an approach similar to ours, but are not identical, especially regarding the exact solution for 
three-flavour oscillations in the cosmological context, with and without including Lindblad operators to 
account for decoherence. In  \cite{Donadi:2013uxa} a spontaneous suppression operator is adopted to account 
for decoherence, but the focus is not the evolution of entropy. Importantly, most existing studies 
are concerned with the interplay of neutrino oscillations and the plasma, therefore their focus lies on the epoch of
neutrino decoupling. To our knowledge, the epoch after neutrino decoupling has not been explored in great detail,
especially the effect of the transition from the relativistic to non-relativistic regime in a statistically 
homogeneous space, which we argue is the source of decoherence proposed in this work.

Often a prescription based on wave packets is adopted to account for the process of decoherence 
\cite{Giunti:2003ax,Akhmedov:2009rb}, e.g.\ for supernova neutrinos 
\cite{PhysRevD.74.105010,roberto,Kersten:2015kio,Akhmedov:2017mcc} or cosmological neutrinos \cite{Pfenniger:2006rd}.
In the latter work it is argued that cosmological neutrinos should not be treated as a classical, collisionless fluid. 
The author of \cite{Bernardini:2012uc} addresses the increase of neutrino entropy 
due to decoherence of mixed massive neutrinos, but without solving the Lindblad equation or  
using wave packets. As we show below, our findings go well beyond the studies presented in \cite{Pfenniger:2006rd} 
and \cite{Bernardini:2012uc}.

Lindblad operators have been used more recently to account for various decoherence processes 
\cite{Adler:2000vfa,Ohlsson:2000mj,2012EJPh33805P,Weinberg:2016uml,Distler:2017kah}, as well as in the 
context of neutrino oscillations in laboratory experiments, and both for two \cite{Ohlsson:2000mj,Oliveira:2014jsa} and for 
three neutrino flavours \cite{Barenboim:2006xt,Gomes:2016ixi,Coelho:2017zes}. However, the use of Lindblad operators for cosmological neutrinos in 
this work is novel, as well as the exact solution for three neutrino flavours that is valid throughout the relativistic and non-relativistic regimes.

Using Lindblad operators, one can mimic the damping of mixed states caused by a hypothetical decoherence phenomenon and predict the net change in entropy in a mathematical rigorous way. We also argue that cosmological neutrinos could be used to improve our understanding of decoherence in analogue to the recently proposed studies of decoherence in atomic clocks \cite{Weinberg:2016uml}.

This work is structured as follows. In the next two sections we present exact solutions for two- and three-flavour neutrino
oscillations in cosmology, including decoherence. We show that suitable Lindblad operators give rise to decoherence and 
that time averaging would lead to the same result for the asymptotic density matrix, we discuss the validity and generality of this result.
In section 4 we study the increase of family entropy associated with decoherence. 
We conclude with a discussion of our findings.

Whenever we use values for neutrino mixing angles and mass square differences, we use the best-fit values
from \cite{Tortola:2012te}: $\sin^2\theta_{12}=0.320^{+0.016}_{-0.017}$, $\sin^2\theta_{23}=0.613^{+0.022}_{-0.040}(0.600^{+0.023}_{-0.031})$\footnote{There is a local minimum for the mixing angle $\theta_{23}$ at $0.427^{0.034}_{0.027}$ with a difference of $\Delta \chi^2 = 0.02$
when compared to the global minimum.}, $\sin^2\theta_{13}=0.0246^{+0.0029}_{-0.0028}(0.0250^{+0.0026}_{-0.0027})$, 
$\Delta m_{21}^2=(7.62\pm 0.19)\times 10^{-5}\text{ eV}^2$ and $| \Delta m_{31}^2 |=2.55^{+0.06}_{-0.09}(2.43^{+0.07}_{-0.06})\times 10^{-3}\text{ eV}^2$. 
Where the values are for normal (inverted) hierarchy. The exception is the Dirac CP-violation phase which we set to
zero for the sake of simplicity, although we explore its effect in the appendix \ref{sec:appendixa}.
The best-fit value for the CP-violation phase is $0.8\pi(-0.03\pi)$, but
the phase is not measured at statically significant level. The convention adopted on the mixing angles follows the standard
of the Particle Data Group \cite{Agashe:2014kda}. The neutrino
oscillation parameters adopted here have been obtained from neutrino
oscillation experiments only \cite{Tortola:2012te} and are
in agreement with an independent recent compilation \cite{Capozzi:2017ipn}, that also uses cosmological information.

\section{Neutrino oscillations: two-flavour case}

We start by a discussion of the two-flavour case. Although some readers might think that this is a trivial exercise, 
we think that it is useful to clarify the concepts and the method for a discussion of the three-flavour case. 
We obtain an exact analytic treatment of neutrino vacuum oscillations from the relativistic to the 
non-relativistic regime in an expanding universe, their decoherence and time 
and/or momentum averaging for arbitrary initial conditions, neutrino parameters and cosmological model parameters. 

The physical state of a system with two neutrino flavours is described by a two-dimensional Hilbert space 
(factored with the corresponding spaces for the other physical degrees of freedom --- neutrino spin and momentum). 
The space of all hermitian $2 \times 2$ matrices is spanned by the unit matrix $I$ and the Pauli matrices $\sigma_i$, where 
the Latin indices $i$ run from $1$ to $3$. We write for any hermitian matrix $M = M_0 I + M_i \sigma_i$, where we sum 
over repeated indices. We have $\mathrm{tr} M = 2 M_0$ and $\mathrm{tr} M^2 = 2 (M_0^2 + M_i^2)$. Note that hermiticity 
implies that the components $M_i$ are real numbers.

{Expressing the density matrix in this matrix basis, the von Neumann equation (\ref{vN}) becomes
\begin{equation}
D \rho_0 = 0, \quad D \rho_i = 2 \epsilon_{ijk} \mathcal{H}_j \rho_k,
\label{vNcomponents}
\end{equation}
with $\epsilon_{ijk}$ denoting the totally antisymmetric symbol. The trace condition gives $\rho_0 = 1/2$.

So far we did not specify a basis for the neutrino states. The free Hamiltonian is diagonal for the neutrino 
mass states and reads
\begin{equation}
\mathcal{H}_{\rm m} = \mathcal{H}_0 I + \mathcal{H}_3 \sigma_3,
\end{equation}
with 
\begin{eqnarray}
\mathcal{H}_0 & =  &  \frac 12 \left( \sqrt{m_1^2 + \frac{q^2}{a^2}} +  \sqrt{m_2^2 + \frac{q^2}{a^2}}\right), \\
\mathcal{H}_3 & =  &  \frac 12 \left( \sqrt{m_1^2 + \frac{q^2}{a^2}}  -  \sqrt{m_2^2 + \frac{q^2}{a^2}}\right).
\end{eqnarray} 
Without restriction of generality we assume that $0 \leq m_1 < m_2$. For $m_1 = m_2$ we find that 
$\mathcal{H}_3 = 0$ and $\rho$ is a constant matrix.  

Flavour mixing is described by a two dimensional rotation ($U^\dagger U = I$), written as 
\begin{equation}
U = \left(\begin{array}{cc} \cos \theta & \sin \theta \\ - \sin \theta & \cos \theta \end{array}\right).
\end{equation}
The Hamiltonian in flavour space reads: 
\begin{equation}
\mathcal{H}_\mathrm{f} = U \mathcal{H}_\mathrm{m} U^\dagger. \label{eq:transtoflavour}
\end{equation} 
Thus the solution in flavour basis is
\begin{equation}
\rho_{\rm f}(a,q) = U \rho_{\rm m}(a,q) U^\dagger, 
\end{equation}
where the initial condition for the mass states is given by the rotated conditions in flavour space
\begin{equation}
\rho_{\rm m}(a_{\rm ini},q) = U^\dagger \rho_{\rm f}(a_{\rm ini},q) U.
\end{equation}

\subsection{Exact solution} 

Neutrino production and detection involves neutrino interactions, i.e.\ flavour states. Thus 
for both the initial conditions and the late time values of $\rho$ we are interested in the flavour basis. Nevertheless, 
the Hamiltonian is diagonal in the mass basis, giving rise to simple time evolution of equation (\ref{vN}). 
We thus first study the time evolution in the mass basis using Bloch vectors. Once the time evolution is known, we can specify the 
initial conditions in flavour basis, transform them to the mass basis, evolve in time and finally transform back to the 
flavour basis. As shown below, this can be done analytically without any assumption on neutrino masses, momenta 
or cosmological model.

In the mass basis the von Neumann equation (\ref{vNcomponents}) is simply,
\begin{equation}
D \rho_0 = 0, \quad D \rho_1 = - 2 \mathcal{H}_3 \rho_2, \quad D \rho_2 = 2 \mathcal{H}_3 \rho_1, \quad 
D \rho_3 = 0.
\end{equation}
Thus $\rho_0$ and $\rho_3$ are constants. Furthermore, we define 
\begin{equation} 
\Delta(a,q) = - 2 \mathcal{H}_3 = \left( \sqrt{m_2^2 + \frac{q^2}{a^2}}  -  \sqrt{m_1^2 + \frac{q^2}{a^2}}\right).
\end{equation}
With the new variable
\begin{equation}
{\rm d} x_q = \frac{\Delta(a,q)}{H(a)} {\rm d} \ln a,   \label{eq:timevariable22}
\end {equation}
we find the exact solution
\begin{equation}
\rho_1(x) = A_q \cos(x_q + \phi_q), \quad \rho_2(x) = - A_q \sin(x_q +\phi_q),
\end{equation}
where $A_q$ and $\phi_q$ are to be fixed by the initial conditions. We find that 
${\rm tr} \rho^2 = 1/2 + 2(\rho_3^2 + A_q^2)$.

As we saw already, the first necessary condition for neutrino oscillation (a non-trivial evolution of the 
density matrix) is $m_2 > m_1$. A second necessary 
condition is $\theta \neq 0$, as for $\theta = 0$ the mass basis agrees with the flavour basis and only non-trivial 
values for $\rho_0$ and $\rho_3$ could be generated (under the assumption that neutrinos can only be 
generated in a pure flavour state). As was shown above, both $\rho_0$ and $\rho_3$ are preserved in the 
mass basis and thus for vanishing mixing angles no neutrino oscillations occur. 

As we show below, there is also a third necessary condition for the oscillations of neutrino flavours to 
happen: at least one of the $\rho_{{\rm f}i} \neq 0$, otherwise the oscillation amplitude $A$ vanishes.  

In order to find explicit expressions we first study how the Pauli matrices are transformed from mass to flavour 
space. The transformation in the other direction is obtained by $\theta \to - \theta$. This allows us to fix the constants 
$A_q$ and $\phi_q$. In the following we drop the explicit indication of the $q$-dependence; we find 
\begin{equation}
A = \sqrt{[\cos(2\theta) \rho_{{\rm f}1}(x_{\rm ini}) + \sin(2\theta) \rho_{{\rm f}3}(x_{\rm ini})]^2 +  \rho_{{\rm f}2}(x_{\rm ini})^2} \label{eq:amplitude22}
\end{equation}
and 
\begin{equation}
\phi = - x_{\rm ini} + \arctan \left(\frac{- \rho_{{\rm f}2}(x_{\rm ini})}{\cos(2\theta) \rho_{{\rm f}1}(x_{\rm ini}) + \sin(2\theta) \rho_{{\rm f}3}(x_{\rm ini})} \right). \label{eq:phase22}
\end{equation}
Finally, at $x > x_{\rm ini}$, we may express the most general solution in flavour space as
\begin{eqnarray}
\rho_{{\rm f}0}(x)  & =  & \frac 12,  \\
\rho_{{\rm f}1}(x)  & =  &  \cos(2\theta) A \cos(x + \phi) + \sin^2(2\theta)  \rho_{{\rm f}1}(x_{\rm ini}) - \sin(2\theta)\cos(2\theta) \rho_{{\rm f}3}(x_{\rm ini}),\\
\rho_{{\rm f}2}(x)  & =  &  - A \sin(x + \phi), \\
\rho_{{\rm f}3}(x)  & =  &  \sin(2\theta) A \cos(x+\phi) - \sin(2\theta) \cos(2\theta) \rho_{{\rm f}1}(x_{\rm ini}) + \cos^2(2\theta) 
\rho_{{\rm f}3}(x_{\rm ini}).
\end{eqnarray} 
It is interesting to check that indeed  
\begin{equation}
\mathrm{tr} \rho^2 =  \frac{1}{2} + 2 [\rho_{{\rm f}1}^2(x_{\rm ini}) + \rho_{{\rm f}2}^2(x_{\rm ini}) + 
\rho_{{\rm f}3}^2(x_{\rm ini})]
\end{equation}
is a preserved quantity. As $\mathrm{tr} \rho^2 \leq 1$ for any physical state, we find that the initial 
conditions have to satisfy the constraint 
\begin{equation}
\rho_{{\rm f}1}^2(x_{\rm ini}) + \rho_{{\rm f}2}^2(x_{\rm ini}) + \rho_{{\rm f}3}^2(x_{\rm ini}) \leq \frac{1}{4}.
\end{equation}
Since the components $\rho_{i}$ are real, it follows that all individual components have to come from the 
interval $[-1/2, + 1/2]$, for any initial conditions including arbitrary lepton-flavour asymmetries. 
Thus we also see that $\mathrm{tr} \rho^2 \geq 1/2$.

For the special case of maximal mixing, given by $\theta = \pi/4$, we find 
\begin{eqnarray}
\rho_{{\rm f}0}(x)  & =  & \frac 12,  \\
\rho_{{\rm f}1}(x)  & =  & \rho_{{\rm f}1}(x_{\rm ini}),  \\
\rho_{{\rm f}2}(x)  & =  &  - A  \sin(x - \phi), \\
\rho_{{\rm f}3}(x)  & =  &   A  \cos(x - \phi).
\end{eqnarray} 
with
\begin{equation}
A = \sqrt{\rho_{{\rm f}3}^2 +  \rho_{{\rm f}2}^2}, \quad \phi = - x_{\rm ini} + \arctan \left(-\frac{\rho_{{\rm f}2}(x_{\rm ini})}{\rho_{{\rm f}3}(x_{\rm ini})} \right).
\end{equation}

\subsection{Initial conditions} 

If at the initial time all neutrinos are in one of the two flavour states, we have
\begin{equation}
\rho_{\rm f}(x_{\rm ini}) = \frac{1}{2}\left(\begin{array}{cc} 1 +\delta & 0 \\ 0 & 1-\delta \end{array}\right) ,
\end{equation}
with $\delta =\delta(q) \in [-1,1]$ describing the asymmetry of the initial flavours. This means we assume 
that nature does not produce flavour-entangled (mixed) states at neutrino decoupling. 
Expanding in Pauli matrices
\begin{equation}
\rho_{{\rm f}0}(x_{\rm ini}) =  \frac 12,\ \rho_{{\rm f}1}(x_{\rm ini})  =  0 ,\ \rho_{{\rm f}2}(x_{\rm ini})  =  0, \ 
\rho_{{\rm f}3}(x_{\rm ini}) =  \frac \delta 2,
\end{equation}
and replacing in (\ref{eq:amplitude22}) and (\ref{eq:phase22}), we have
\begin{equation}
A = \frac{\delta}{2}\sin(2\theta) , \quad \phi = - x_{\rm ini}, 
\end{equation}
and finally
\begin{eqnarray}
\rho_{{\rm f}0}(x)  & =  & \frac 1 2,  \label{eq:solution22flavour0}\\
\rho_{{\rm f}1}(x)  & =  &  \frac{\delta}{2}[\sin(2\theta) \cos(2\theta)  \cos(x - x_{\rm ini}) -  \sin(2\theta)\cos(2\theta)], \label{eq:solution22flavour1}\\
\rho_{{\rm f}2}(x)  & =  &  - \frac{\delta}{2}\sin(2\theta)  \sin(x - x_{\rm ini}) , \label{eq:solution22flavour2} \\
\rho_{{\rm f}3}(x)  & =  &  \frac{\delta}{2}[\sin^2(2\theta) \cos(x - x_{\rm ini}) + \cos^2(2\theta)] . \label{eq:solution22flavour3}
\end{eqnarray} 
In the following we choose $x_{\rm ini} = 0$, without restriction of generality.

\subsection{Decoherence}

To describe the decoherence of a system of two neutrino flavours in vacuum in an expanding Universe, 
we make use of a single Lindblad operator \cite{Ohlsson:2000mj} and decompose it in Pauli matrices
\begin{equation}
{\cal L} = l_0\boldsymbol{I} + \sum_i l_i\sigma_i \ ,
\end{equation}
where we have in principle four amplitudes $l_i$. Any Lindblad operator ${\cal L}_a$ has to be hermitian 
(${\cal L}^\dag ={\cal L}$) to guarantee a monotonic increase of the von Neumann entropy and it has to commute 
with the Hamiltonian ($[{\cal H}_m,{\cal L}]=0$) in order to conserve the statistical average energy 
$\left(\frac{d}{dt}{\rm tr}({\cal H}_m\rho^{\rm L})=0\right)$. Commutation with the diagonal 
Hamiltonian (in the mass basis in vacuum) requires that $l_1=l_2=0$, resulting in the decoherence operator
\begin{equation}
 \left[{\cal L},\left[\rho^{\rm L},{\cal L}\right]\right] = -4l_3^2 (\sigma_1\rho_1^{\rm L}+\sigma_2\rho_2^{\rm L}) \ .
\end{equation}
Without restriction of generality we put $l_0=0$, as it does not show up in the Lindblad equation.
Thus, the components of the master equation with the Lindblad operator ${\cal L}$ become
\begin{equation}
D \rho^{\rm L}_0 = 0, \quad D \rho^{\rm L}_1 = - 2 \mathcal{H}_3 \rho^{\rm L}_2 - 4l_3^2 \rho^{\rm L}_1, \quad 
D \rho^{\rm L}_2 = 2 \mathcal{H}_3 \rho^{\rm L}_1 - 4l_3^2 \rho^{\rm L}_2, \quad D \rho^{\rm L}_3 = 0,
\end{equation}
whose solution in mass basis can also be found analytically (using $d t = - d x/2 {\cal H}_3$),
\begin{eqnarray}
\rho^{\rm L}_{0}(x)  &=&  \frac 12, \label{eq:sollindblad0} \\
\rho^{\rm L}_{1}(x)  &=&  
\frac \delta 2  \sin (2\theta) \cos (x)\exp\left[{2\int \frac{l_3^2}{\mathcal{H}_3} dx}\right], \label{eq:sollindblad1} \\
\rho^{\rm L}_{2}(x)  &=&    
- \frac \delta 2 \sin(2\theta) \sin(x)\exp\left[{2\int \frac{l_3^2}{\mathcal{H}_3} dx}\right], \label{eq:sollindblad2} \\
\rho^{\rm L}_{3}(x)  &=&   \frac \delta 2  \cos(2\theta), \label{eq:sollindblad3}
\end{eqnarray} 
and are easily rotated to the flavour basis
\begin{eqnarray}
\rho^{\rm L}_{{\rm f}0}(x)  &=&  \frac 1 2,  \label{eq:sollindbladflavour0} \\
\rho^{\rm L}_{{\rm f}1}(x)  &=&  
 -\frac \delta 2  \sin (2\theta) \cos (2 \theta)\left(1 - \cos (x)\exp\left[{2\int \frac{l_3^2}{\mathcal{H}_3} dx}\right]\right), \label{eq:sollindbladflavour1} \\
\rho^{\rm L}_{{\rm f}2}(x)  &=&    - \frac \delta 2 \sin(2\theta) \sin(x)\exp\left[{2\int \frac{l_3^2}{\mathcal{H}_3} dx}\right], \label{eq:sollindbladflavour2} \\
\rho^{\rm L}_{{\rm f}3}(x)  &=&  
\frac \delta 2  \left(\cos^2(2\theta) + \sin^2(2 \theta)\cos (x)\exp\left[{2\int \frac{l_3^2}{\mathcal{H}_3} dx}
\right]\right). \label{eq:sollindbladflavour3}
\end{eqnarray} 
Note that $l_3$ can be an arbitrary real function of $x_q$. The integral in the exponent is negative definite if 
$l_3 \neq 0$ and thus gives rise to a damping of all non-diagonal components in the mass basis. 
We interpret the function $l_3(x)$ as the influence of the environment, the expanding cosmos,
that leads to the decoherence of neutrino states.}

\subsection{Averaging}

The time averaging of (\ref{eq:solution22flavour0}) -- (\ref{eq:solution22flavour3}) produces 
the same asymptotic result as the decoherence process by the Lindblad operator. 
The final effect is the suppression of terms with time dependence. 
For time averaging the fast oscillations terms take asymptotic values ($\langle \cos(x)\rangle = 
\langle \sin(x) \rangle \rightarrow 0$ and $\langle \sin^2(x) \rangle =\langle \cos^2(x)\rangle \rightarrow 1/2$). 
Using Lindblad operators, one can see that the terms which have a time dependence in the mass basis, equations
(\ref{eq:sollindblad1}) and (\ref{eq:sollindblad2}), are exponentially suppressed once decoherence starts, 
extinguishing the off-diagonal contributions corresponding to mixed states. In the flavour basis, the off-diagonal terms 
are driven to a constant value. We use the effect of the Lindblad operator to describe decoherence, setting to zero 
the contributions from $\rho_1$ and $\rho_2$ in the mass basis and then rotate to the flavour basis
\begin{eqnarray}
\bar \rho_{{\rm f}0}  & =  & \frac 1 2,  \label{eq:averagedflavour0} \\ 
\bar \rho_{{\rm f}1}  & =  &  -\sin(2\theta)\cos(2\theta) \frac \delta 2,  \label{eq:averagedflavour1} \\
\bar \rho_{{\rm f}2}  & =  &  0, \label{eq:averagedflavour2} \\
\bar \rho_{{\rm f}3}  & =  &  \cos^2(2\theta) \frac \delta 2. \label{eq:averagedflavour3}
\end{eqnarray} 

For this averaged system we have $\mathrm{tr}\bar\rho^2 = \frac{1}{2}[1 + \delta^2 \cos^2(2\theta)]$, which is independent 
of time or oscillation phase since we performed averaging and it depends on the mixing angle because it 
undergoes mixing. 
The result is numerically equivalent to a system that lost all coherence. 
For the exact, non-averaged solution we have $\mathrm{tr}\rho^2 = \frac{1}{2}(1 + \delta^2)$, 
which is independent of mixing angles and time, as one expects for a system that undergoes unitary
(deterministic) time evolution.
The difference of the traces of the squared matrices, a measure for the difference of coherence of averaged and microscopic states, is 
then $\mathrm{tr}\bar\rho^2 - \mathrm{tr}\rho^2 = - \frac{1}{2}\delta^2 \sin^2(2\theta)$. 

For maximal mixing ($\theta=\pi/4$), the time averaged solution with arbitrary initial condition becomes
\begin{equation}
\bar \rho_{{\rm f}0}  =   1/2,  \ \bar \rho_{{\rm f}1}   =   \rho_{{\rm f}1} (x_{\rm ini}), \ 
\bar \rho_{{\rm f}2}  =  0, \ \bar \rho_{{\rm f}3}  =    0.
\end{equation} 
Thus the density matrix of maximally mixing neutrinos does not depend on the mixing of the initial flavour distortion. 
The probabilities to find a neutrino in the first or second flavour state are equal, and the amount of mixing is constant.
For initial conditions in which the neutrinos are in pure flavour states ($\rho_{{\rm f}1} (x_i)=0$), 
the time averaged density matrix is proportional to the unit matrix, as one would expect.

\subsection{Relation of Lindblad formalism and averages}

We observe that the expressions (\ref{eq:averagedflavour0}) -- (\ref{eq:averagedflavour3}) are identical
to the expressions (\ref{eq:sollindbladflavour0}) -- (\ref{eq:sollindbladflavour3}) asymptotically ($\rho^{\rm L}_{{\rm f}i}
\simeq \bar \rho_{{\rm f}i}$), i.e.\ if the oscillation phase $x$ is large enough then decoherence has happened. 
The averaged density matrix agrees with the microscopic density matrix after decoherence.  It might be tempting to 
conclude that decoherence and averaging over time (or perhaps momenta) would be equivalent. This has actually 
be proposed in previous literature, see e.g.~\cite{Ohlsson:2000mj}. However, a closer investigation reveals that this is not the case.

Let us apply our calculation to experiments which test solar, reactor or 
atmospheric neutrinos, in order to revisit the arguments given in \cite{Ohlsson:2000mj}. 
The neutrinos of interest in oscillation experiments are relativistic, i.e.\ $q \approx E$ and thus 
$\Delta(a,q) \approx \Delta m^2/2E$ (with energy $E$),  and propagate in non-expanding space 
[$(1/H) {\rm d} \ln a = dt, a = 1$]. We find the oscillation phase (\ref{eq:timevariable22}) becomes
\begin{equation}
  x_q = \int \frac{\Delta(a,q)}{H(a)} {\rm d} \ln a \approx \int \frac{\Delta m^2}{2E} dt = \frac{\Delta m^2}{2E} L, 
\end{equation}
where the distance $L$ is the distance travelled by neutrinos at the speed of light. 
The decoherence term becomes
\begin{equation}
\exp\left(4\int_0^{L/c} \frac{l_3^2}{2 \mathcal{H}_3} dx\right) = \exp(-4\int_0^{L/c} \!\!\! l_3^2 dt) \ .
\end{equation}
For the analysis of neutrino oscillation experiments, one usually measures the amount of neutrinos that 
were emitted in one flavour and are detected in either the same or the other flavour. This corresponds to a 
maximal initial distortion (e.g. $\delta = 1$ corresponds to $\rho(x_{\rm ini}) = \left| \nu_1 \rangle \langle \nu_1 \right|$). 
The well-known result, including a decoherence term, for the probability to measure the second flavour is now 
easily recovered, 
$P_{1 \to 2}(L,E) = \mbox{tr}[\left| \nu_2 \rangle \langle \nu_2 \right| \rho(x)]$)
\begin{equation}
 P_{1 \to 2}(L,E) = \frac{1}{2}\sin^2(2\theta)\left[1-\exp({-4\int_0^{L/c} \!\!\! l_3^2 dt})\cos\left(\frac{\Delta m^2}{2E} L \right)  \right] \ . 
 \label{eq:oscilllindblad}
\end{equation}

Let us compare this result with the probability obtained in a description in which wave packets instead of 
Lindblad operators (i.e. $l_3 = 0$) are considered to describe the process of decoherence 
\cite{Ohlsson:2000mj,Giunti:2003ax}. The distribution of the oscillation phase $x$ is often assumed to be a Gaussian, 
the phase averaged probability of flavour oscillations is then given by
\begin{equation}
\label{eq_wp}
 \left\langle P_{1 \to 2}  \right\rangle = \int P_{1 \to 2}(x) \left[ \frac{1}{\sigma\sqrt{2\pi}}\exp\left( -\frac{(x-\left\langle x \right\rangle)^2}{2\sigma^2} \right) \right] dx \ ,
\end{equation}
where the phase width of the wave packet is $\sigma=\sqrt{\left\langle (x-\left\langle x \right\rangle)^2 \right\rangle}$. 
The integral gives
\begin{equation}
\label{eq_gaussianaverage}
\langle P_{1 \to 2} \rangle = 
\frac{1}{2}\sin^2(2\theta)\left(1-\exp\left(-\frac{\sigma^2}{2}\right) \cos(\langle x \rangle) \right) \ ,
\end{equation}
with $\langle x \rangle = (\Delta m^2/2)  \langle L/E \rangle.$ 
Comparing the equation above with equation (\ref{eq:oscilllindblad}), it has been argued that they have the same 
structure once we identify the decoherence term with the wave packet dispersion
\begin{equation}
\label{eq_comparison}
4 \int^{L/c}_{0} l_3^2 d t = \frac{\sigma^2}{2}.
\end{equation}
This result is the basis of the argued equivalence between Lindblad decoherence and Gaussian averaging 
\cite{Ohlsson:2000mj}. 

However, this argument is inconsistent already at the level of mathematical assumptions. In order to go from 
(\ref{eq_wp}) to (\ref{eq_gaussianaverage}), $\sigma$ was assumed to be a constant, i.e. independent of 
the phase $x$. But the identification (\ref{eq_comparison}) suggests that either $\sigma = \sigma(L)$, inconsistent with the 
assumption, or the integral on the l.h.s.\ should be constant. To make that integral a constant we can assume that $l_3^2$ 
is proportional to a Dirac-delta distribution on the time interval $[0,L/c]$, which would correspond to a sudden, but 
incomplete decoherence, i.e. the asymptotic state remains to show (damped) neutrino oscillations. We conclude that the two mechanisms, decoherence and phase (or time or momentum) averaging are not physically equivalent.  

\section{Neutrino oscillations: three-flavour case}

For the three flavour case the space of hermitian matrices is spanned by the unit matrix and the Gell-Mann matrices 
$\lambda_i$ \cite{Arfken} where the index $i$ runs from 1 to 8. In this case, equation (\ref{vN}) becomes
\begin{equation}
D \rho_0 = D \rho_3 = D \rho_8 = 0, \quad D \rho_k = 2 f_{i3k} \rho_i \mathcal{H}_3 + 2 f_{i8k} \rho_i \mathcal{H}_8 \ ,
\label{vNcomponentsthree}
\end{equation}
where $f_{ijk}$ are the usual structure constants of the Lie algebra su(3).
By unitarity, we mandatorily have $\rho_0 = 1/3$. 

In the mass basis the Hamiltonian is diagonal and is given by
\begin{equation}
\mathcal{H}_{\rm m} = \mathcal{H}_0 I + \mathcal{H}_3 \lambda_3 + \mathcal{H}_8 \lambda_8 ,
\end{equation}
with 
\begin{eqnarray}
\mathcal{H}_0 & =  &  \frac{1}{3} \left( \sqrt{m_1^2 + \frac{q^2}{a^2}} +  \sqrt{m_2^2 + \frac{q^2}{a^2}} +  \sqrt{m_3^2 + \frac{q^2}{a^2}} \right), \\
\mathcal{H}_3 & =  &  \frac{1}{2} \left( \sqrt{m_1^2 + \frac{q^2}{a^2}}  -  \sqrt{m_2^2 + \frac{q^2}{a^2}}\right), \\
\mathcal{H}_8 & =  &  \frac{1}{2\sqrt{3}} \left( \sqrt{m_1^2 + \frac{q^2}{a^2}}  + \sqrt{m_2^2 + \frac{q^2}{a^2}} -2\sqrt{m_3^2 + \frac{q^2}{a^2}}\right).
\end{eqnarray} 
Without restriction of generality we assume that $0 \leq m_1 < m_2 < m_3$ for normal hierarchy and $0 \leq m_3 < m_1 < m_2$ for inverted hierarchy. Once again, for equal or vanishing masses 
the Hamiltonian is proportional to the identity matrix. 

The flavour mixing matrix is now a three dimensional rotation ($U^\dagger U = I$) and can be written as 
{\small
\begin{equation}
 U = \left(       \begin{array}{ccc}
c_{12}c_{13} 				& s_{12}c_{13}   			& s_{13}e^{-i\delta_{\rm CP}}  \\
-s_{12}c_{23}-c_{12}s_{23}s_{13}e^{i\delta_{\rm CP}}  	& c_{12}c_{23}-s_{12}s_{23}s_{13}e^{i\delta_{\rm CP}} 	& s_{23}c_{13} \\
s_{12}s_{23}-c_{12}c_{23}s_{13}e^{i\delta_{\rm CP}}  	& -c_{12}s_{23}-s_{12}c_{23}s_{13}e^{i\delta_{\rm CP}}   	& c_{23}c_{13}
											\end{array}
         \right),
\end{equation}
}
where $c_{ij}=\cos\theta_{ij}$, $s_{ij}=\sin\theta_{ij}$ and $\delta_{\rm CP}$ is a Dirac charge-parity (CP) violating phase. 
In the main body of this work we assume vanishing CP-violation and include some results for a non-vanishing value
in the appendix \ref{sec:appendixa}. Both Majorana phases are irrelevant for the aspects discussed in this work.
The Hamiltonian in flavour space is obtained as in equation (\ref{eq:transtoflavour}).

\subsection{Exact solution}

While the initial states and the states observable by means of a particle physics detector are given in flavour basis, 
the von Neumann equation and its solution is most suitable formulated in the mass basis. This approach simplifies the 
system of equations since the Hamiltonian is diagonal in the mass basis,
\begin{eqnarray}
& D \rho_0 = 0 , \ D \rho_1 = -2\mathcal{H}_3\rho_2 , \ D \rho_2 = 2\mathcal{H}_3\rho_1 , \nonumber \\ 
& D \rho_3 = 0 , \ D \rho_4 = -\left(\mathcal{H}_3+\sqrt{3}\mathcal{H}_8 \right)\rho_5 , \ D \rho_5 = \left(\mathcal{H}_3+\sqrt{3}\mathcal{H}_8 \right)\rho_4 ,  \\
& D \rho_6 = -\left(-\mathcal{H}_3+\sqrt{3}\mathcal{H}_8 \right)\rho_7 , \ D \rho_7 = \left(-\mathcal{H}_3+\sqrt{3}\mathcal{H}_8 \right)\rho_6 , \ D \rho_8 = 0 . \nonumber
\label{vNcomponentsthreen}
\end{eqnarray}
The diagonal components of the density matrix ($\rho_0,\rho_3$ and $\rho_8$) are constant. The remaining 
off-diagonal components give rise to oscillating solutions. The oscillation frequency is determined by   
combinations of the asymmetric terms of the Hamiltonian ($\mathcal{H}_3$ and $\mathcal{H}_8$). 
There are six equations, forming three independent harmonic oscillators of two levels each, where their frequency is given by
\begin{equation} 
\Delta(a,q)_{21} = - 2 \mathcal{H}_3 = \left( \sqrt{m_2^2 + \frac{q^2}{a^2}}  -  \sqrt{m_1^2 + \frac{q^2}{a^2}}\right) \ ,
\end{equation}
\begin{equation} 
\Delta(a,q)_{31} = - (\mathcal{H}_3+\sqrt{3}\mathcal{H}_8) = \left( \sqrt{m_3^2 + \frac{q^2}{a^2}}  -  \sqrt{m_1^2 + \frac{q^2}{a^2}}\right) \ ,
\end{equation}
\begin{equation} 
\Delta(a,q)_{32} = - (-\mathcal{H}_3+\sqrt{3}\mathcal{H}_8) = \left( \sqrt{m_3^2 + \frac{q^2}{a^2}}  -  \sqrt{m_2^2 + \frac{q^2}{a^2}}\right) \ ,
\end{equation}
for simplicity, we define three different oscillation phases to account for the three sectors of oscillation
\begin{equation}
{\rm d} x_{ij} = \frac{\Delta(a,q)_{ij}}{H(a)} {\rm d} \ln a\ ,
\end {equation}
where the only three independent combinations are $ij=21,31,32$. We find the exact solutions
\begin{eqnarray}
\rho_{1}(x_{21}) = + A \cos(x_{21} + \phi_{{12}}), \quad \rho_{2}(x_{21}) = - A \sin(x_{21} + \phi_{{12}}), \nonumber \\
\rho_{4}(x_{31}) = + B \cos(x_{31} + \phi_{{45}}), \quad \rho_{5}(x_{31}) = - B \sin(x_{31} + \phi_{{45}}), \nonumber \\
\rho_{6}(x_{32}) = + C \cos(x_{32} + \phi_{{67}}), \quad \rho_{7}(x_{32}) = - C \sin(x_{32} + \phi_{{67}}).
\end{eqnarray}
The amplitudes $A$, $B$ and $C$ and phases $\phi_{{12}}$, $\phi_{{45}}$ and $\phi_{{67}}$ 
are fixed by the initial conditions. For arbitrary initial conditions, we find 
${\rm tr} \rho^2 = 1/3 + 2(\rho_3^2 + \rho_8^2 + A^2 + B^2 + C^2)$. 

\begin{figure}
\includegraphics[width=0.5\textwidth]{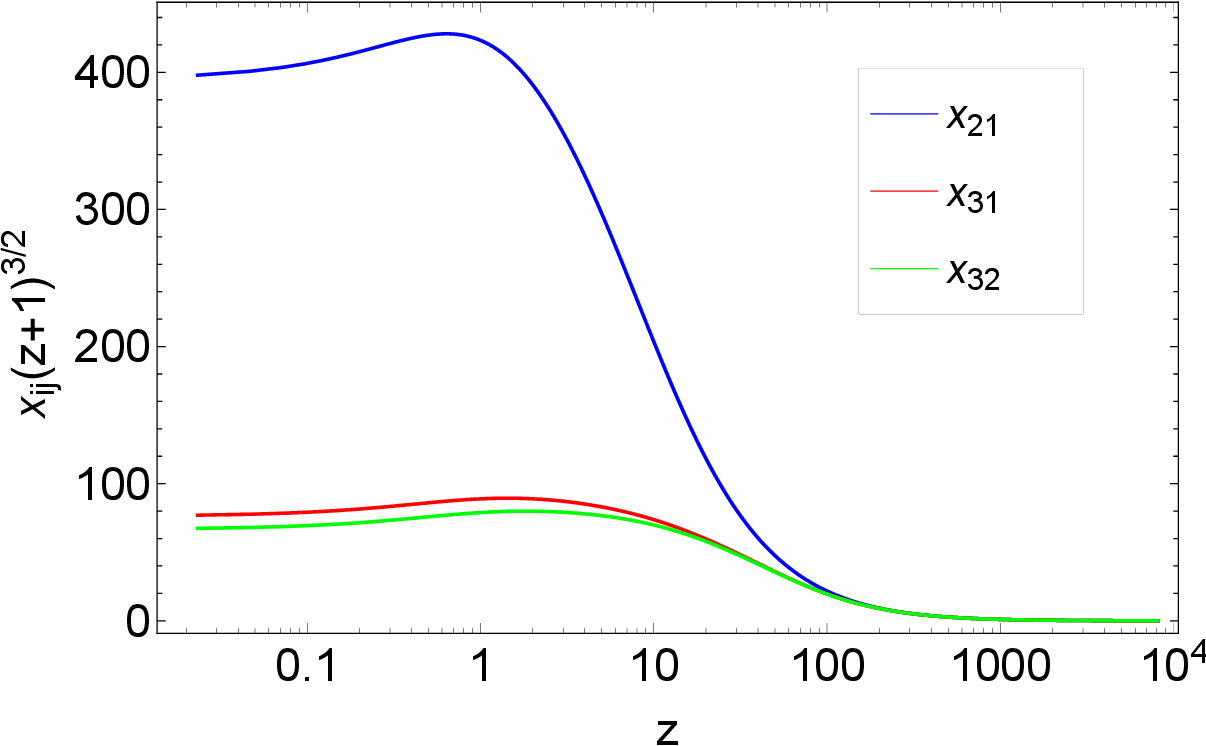}
\includegraphics[width=0.5\textwidth]{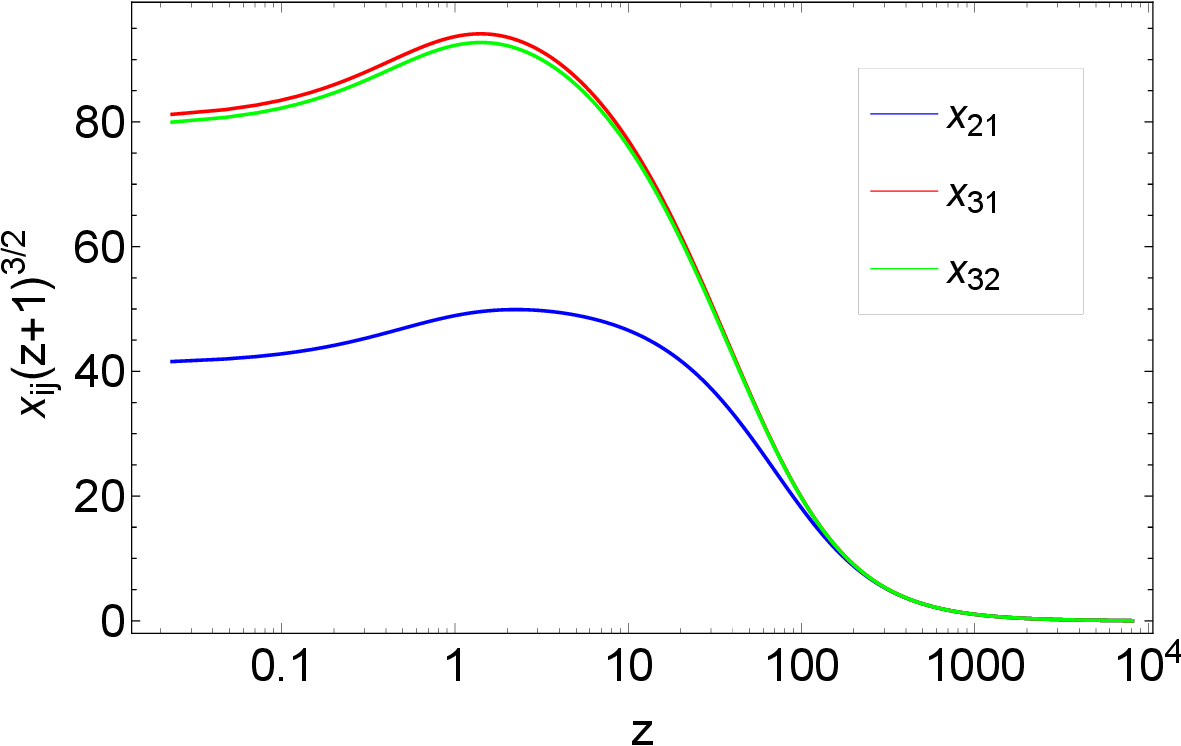}
\caption{Evolution of the neutrino oscillation phases ($x_{21}$, $x_{31}$ and $x_{32}$) as a function of cosmological 
redshift $z$ for the three Gell-Mann blocks for normal (left) and inverted (right) hierarchy with the lightest mass state 
assumed to be massless, shown for neutrinos at the peak of their thermal distribution
($q = 3.15T_\nu$). The mass-squared differences are inferred from the global analysis of neutrino oscillation 
data \cite{Tortola:2012te}.}
\label{graphentropyn}	
\end{figure}

During the radiation dominated epoch all three neutrinos are relativistic and the oscillation phase can be approximated 
by $x_{ij} \approx 1/(\sqrt{\Omega_{\rm rad}} H_0) \Delta m_{ij}^2/[q (1+z)^2]$. When the neutrinos become non-relativistic, 
the oscillation phases start to evolve differently with redshift. In figure \ref{graphentropyn} we show how the three 
different oscillation phases start to deviate from the phase evolution of relativistic neutrinos during the matter dominated 
epoch. Thus we compare to the redshift dependence of the relativistic case, 
$x_{ij} \approx 1/(\sqrt{\Omega_{\rm m}} H_0) \ \Delta m_{ij}^2/[3 q (1+z)^{3/2}]$. As can be clearly seen in the figure, at the 
moment when the most massive neutrino becomes non-relativistic, which happens at $z \sim 100$ for the cases considered, 
the phases start to evolve quite differently, until they evolve again in parallel in the non-relativistic regime.
It is worth noting that the heavier the neutrinos are, the earlier the transition to the non-relativistic regime happens. 
In figure~\ref{graphentropyn} we adopted the lightest mass state to be massless. Thus, one can infer that the transition 
happens at redshifts $z \gtrsim 100$. 

The Gell-Mann matrices are transformed from mass to flavour space according to $U \lambda_i U^\dagger$. 
Instead of presenting the most general solution, we restrict our attention to initial states relevant to cosmology. 

\subsection{Initial conditions}

Just before neutrino decoupling muon and tau neutrinos interact via neutral currents only, while electron neutrinos 
also experience charge current interactions. Thus the muon and tau neutrinos are expected to decouple slightly 
before the electron neutrinos \cite{Mangano:2001iu,Mangano:2005cc}. 
Neutrino oscillations started in the early Universe slightly before their decoupling (defined as the moment when the 
interaction rate equals the Hubble expansion rate) from the primordial plasma. 
Non-instantaneous decoupling and quantum electro-dynamical corrections lead to spectral distortions. 
At this stage, the universe was filled with free electrons and positrons but not anymore with muons or taus, 
which had long annihilated or decayed to lighter leptons or photons. 
Thus electron neutrinos end up with a different distortion of momentum and total number density than the other two 
active flavours. Later, during electron-positron annihilation, extra distortions are produced and again with different 
branching ratios for electron neutrinos and muon/tau neutrinos.

We assume that electron neutrinos are created with a distortion in the density matrix denoted by $\delta(q)$ and 
that muon and tau neutrinos can be described by the same distortion,  $-\delta(q)/2$. We further allow for a difference
in the spectral distortion of muon and tau neutrinos described by $\beta=\beta(q)$. Without primordial lepton-flavour
asymmetry we would expect that $\beta=0$.

Without loss of generality  we start to count oscillations at the moment when we set the initial conditions and thus have 
$x_{21}(t_{\rm ini}) = x_{31}(t_{\rm ini}) = x_{32}(t_{\rm ini}) = 0$. The assumption of pure initial flavour states gives then 
$\phi_{21} = \phi_{31} = \phi_{32} = 0$. The initial conditions are
\begin{equation}
\rho_{\rm f}(0) = \frac{1}{3}\left(\begin{array}{ccc} 1 +\delta & 0 & 0 \\ 0 & 1-\delta/2 +\beta& 0 \\ 0 & 0 & 1-\delta/2  - \beta \end{array}\right) , \label{eq:initialthreeflavours}
\end{equation}
with $\delta \in [-1,2], \beta \in [-\alpha,\alpha]$, with $\alpha = \sqrt{3 - 3 \delta^2/4}$ from the condition 
$\mbox{tr}\rho^2 \leq 1$.
Setting the off-diagonal initial condition null is equivalent to assume that nature does not produce flavour entangled states, 
which is an approximation since neutrinos decoupling is not instantaneous. 

Using the above initial conditions in the equations (\ref{vNcomponentsthreen}), we are left with the following 
initial values for the density matrix in flavour basis (spanned by Gell-Mann matrices)
\begin{equation}
\rho_{{\rm f}0}(0)  =  \frac{1}{3},  \quad \rho_{{\rm f}3}(0)  =  \frac{1}{12} (3 \delta -2 \beta ), \quad \rho_{{\rm f}8}(0)  =  \frac{1}{4\sqrt{3}} \left(\delta + 2 \beta \right),
\end{equation} 
where unshown entries are null. Rotating to the mass basis, we obtain an exact solution of the 
von Neumann equation for the Gell-Mann coefficients of the density matrix,
\begin{eqnarray}
&&\rho_{{\rm }0}  =  \frac{1}{3} ,  \label{eq:initerm0} \\
&&\rho_{{\rm }1}(x_{21})   =  \frac{\delta}{4} \sin (2 \theta_{12}) \cos^2 \theta_{13}\cos (x_{21})\nonumber \\
&&+\frac{\beta}{12}\left[\sin (2 \theta_{12}) (\cos (2 \theta_{13})-3) \cos (2\theta_{23})-4\cos (2 \theta_{12}) \sin \theta_{13} \sin (2 \theta_{23})\right]\cos (x_{21}), \label{eq:soltwo1} \\
&&\rho_{{\rm }2}(x_{21})  =  -\frac{1}{4} \delta  \sin (2 \theta_{12}) \cos^2 \theta_{13}\sin (x_{21})\nonumber \\
&&-\frac{\beta}{12} \left[\sin (2 \theta_{12}) (\cos (2 \theta_{13})-3) \cos (2 \theta_{23}) - 4\cos (2 \theta_{12}) \sin \theta_{13} \sin (2 \theta_{23})\right]\sin (x_{21}), \label{eq:soltwo2} \\
&&\rho_{{\rm }3}  =  \frac{\delta}{4} \cos(2 \theta_{12}) \cos^2\theta_{13} \nonumber \\
&& + \frac{\beta}{12} \left[\cos (2 \theta_{12}) (\cos (2 \theta_{13})-3) \cos (2 \theta_{23}) + 4 \sin (2 \theta_{12}) \sin \theta_{13} \sin (2 \theta_{23})\right], \\  \label{eq:initerm3} 
&&\rho_{{\rm }4}(x_{31})  =  \frac{\delta}{4}   \cos \theta_{12} \sin (2 \theta_{13}) \cos (x_{31}) \nonumber \\
&&+\frac{\beta}{6} \left[\cos \theta_{12} \sin (2 \theta_{13}) \cos (2 \theta_{23}) -2\sin \theta_{12} \cos \theta_{13} \sin (2 \theta_{23})\right] \cos (x_{31}), \label{eq:soltwo4} \\
&&\rho_{{\rm }5}(x_{31})  =  -\frac{\delta}{4} \cos \theta_{12} \sin (2\theta_{13}) \sin (x_{31})\nonumber \\
&& -\frac{\beta}{6}\left[ \cos \theta_{12} \sin (2\theta_{13}) \cos (2 \theta_{23})  -2 \sin \theta_{12}\cos \theta_{13} \sin (2 \theta_{23})\right] \sin (x_{31}) , \label{eq:soltwo5} \\
&&\rho_{{\rm }6}(x_{32})  =  +\frac{\delta}{4} \sin \theta_{12} \sin (2\theta_{13}) \cos (x_{32})\nonumber\\
&&+\frac{\beta}{6}\left[ \sin \theta_{12} \sin (2\theta_{13}) \cos (2 \theta_{23}) +2\cos \theta_{12}\cos \theta_{13} \sin (2\theta_{23})\right] \cos (x_{32}), \label{eq:soltwo6} \\
&&\rho_{{\rm }7}(x_{32})  =  -\frac{\delta}{4} \sin \theta_{12} \sin (2\theta_{13}) \sin (x_{32}) \nonumber \\
&& -\frac{\beta}{6} \left[\sin \theta_{12} \sin (2\theta_{13}) \cos (2 \theta_{23})   +2 \cos \theta_{12}\cos \theta_{13} \sin (2 \theta_{23})\right] \sin (x_{32}), \label{eq:soltwo7} \\
&&\rho_{{\rm }8}  =  +\frac{\delta}{8 \sqrt{3}}  \left( 3\cos(2 \theta_{13})-1\right) +\frac{\beta}{2 \sqrt{3}} \cos^2(\theta_{13}) \cos (2 \theta_{23}) \label{eq:initerm8}  , 
\end{eqnarray}
where the coefficients $\rho_{{\rm }0}$, $\rho_{{\rm }3}$ and $\rho_{{\rm }8}$ are time-independent. 
The corresponding result including non-vanishing CP-violation phase is shown in the appendix.

We do not present the general expressions in the flavour basis because they are lengthy and for the purpose of 
calculating solutions after decoherence, the solution in mass basis is all we need. Instead we restrict our 
presentation to the case of the normal neutrino hierarchy with vanishing Dirac CP-violation phase and for the 
best-fit values of the mixing angles from neutrino oscillation data,
\begin{eqnarray}
\rho_{{\rm f}0}(x_{ij})  & =  & 1/3,  \\
\rho_{{\rm f}1}(x_{ij})  & =  & \delta[-0.08223 + 0.02458\cos (x_{21}) + 0.03569\cos (x_{31}) + 0.02198\cos (x_{32})]+ \nonumber\\ 
                              &&\beta[-0.02451 + 0.00191\cos (x_{21}) - 0.10673\cos (x_{31}) + 0.12933\cos (x_{32})], \\	                        
\rho_{{\rm f}2}(x_{ij})  & =  & \delta[-0.13978\sin (x_{21}) - 0.04476\sin (x_{31}) - 0.01588\sin (x_{32})]+ \nonumber\\ 
                              &&\beta[-0.01086\sin (x_{21}) + 0.13389\sin (x_{31}) - 0.09345\sin (x_{32})], \\
\rho_{{\rm f}3}(x_{ij})  & =  & \delta[0.08141 + 0.14924\cos (x_{21}) + 0.03054\cos (x_{31}) -0.01119\cos (x_{32})]+ \nonumber\\ 
                              &&\beta[-0.02108+ 0.01159\cos (x_{21}) - 0.09134\cos (x_{31}) -0.06584\cos (x_{32})], \\
\rho_{{\rm f}4}(x_{ij})  & =  & \delta[0.03797 - 0.08378\cos (x_{21}) + 0.03559\cos (x_{31}) + 0.01022\cos (x_{32})]+ \nonumber\\ 
                              &&\beta[0.05279 - 0.00651\cos (x_{21}) - 0.10645\cos (x_{31}) + 0.06016\cos (x_{32})], \\
\rho_{{\rm f}5}(x_{ij})  & =  & \delta[0.17592\sin (x_{21}) - 0.02833\sin (x_{31}) - 0.01986\sin (x_{32})]+ \nonumber\\ 
                              &&\beta[0.01366\sin (x_{21}) + 0.08473\sin (x_{31}) - 0.11685\sin (x_{32})], \\
\rho_{{\rm f}6}(x_{ij})  & =  & \delta[-0.09714 + 0.10882\cos (x_{21}) - 0.00012\cos (x_{31}) - 0.01180\cos (x_{32})]+ \nonumber\\ 
                              &&\beta[0.06137  + 0.00845\cos (x_{21}) - 0.00035\cos (x_{31}) - 0.06947\cos (x_{32})], \\
\rho_{{\rm f}7}(x_{ij})  & =  & \delta[-0.03569\sin (x_{21}) + 0.03568\sin (x_{31})  - 0.03568\sin (x_{32})]+ \nonumber\\ 
                              &&\beta[-0.00277\sin (x_{21}) - 0.10673\sin (x_{31})  - 0.20999\sin (x_{32})], \\
\rho_{{\rm f}8}(x_{ij})  & =  & \delta[0.03621  + 0.10009\cos (x_{21}) - 0.02463\cos (x_{31}) + 0.03268\cos (x_{32})]+ \nonumber\\ 
                              &&\beta[0.01492  + 0.00777\cos (x_{21}) + 0.07368\cos (x_{31}) + 0.19230\cos (x_{32})].
\end{eqnarray} 

\subsection{Decoherence}

Similar to the two-flavour case, the Lindblad operator for the three-flavour case has contributions from the 
the same basis elements as the Hamiltonian ($[{\cal H}_{\rm m},{\cal L}]=0$), therefore its form in Gell-Mann matrices is
\begin{equation}
{\cal L} = l_0 I + l_3 \lambda_3 + l_8 \lambda_8.
\end{equation}
Consequently, the decoherence term reads
\begin{eqnarray}
 [{\cal L},[\rho^{\rm L},{\cal L}]] = -4l_3^2(\rho_1^{\rm L}\lambda_1 +\rho_2^{\rm L}\lambda_2) - (l_3+\sqrt{3}l_8)^2(\rho_4^{\rm L}\lambda_4+\rho_5^{\rm L}\lambda_5) \nonumber \\ - (l_3-\sqrt{3}l_8)^2(\rho_6^{\rm L}\lambda_6+\rho_7^{\rm L}\lambda_7).
\end{eqnarray}
Apparently, there is no contribution of the Lindblad term to the evolution equations of $\rho_0, \rho_3$ and 
$\rho_8$, which thus remain constant in the mass basis. As above, we set $l_0=0$.
The Lindblad equation (\ref{vNcomponentsthreen}) can be written as
\begin{eqnarray}
D \rho_0^{\rm L} &=& 0 , \nonumber \\
D \rho_1^{\rm L} &=& -2\mathcal{H}_3\rho_2^{\rm L}  -4l_3^2\rho_1^{\rm L}, \nonumber \\
D \rho_2^{\rm L} &=& 2\mathcal{H}_3\rho_1^{\rm L} -4l_3^2\rho_2^{\rm L}, \nonumber \\
D \rho_3^{\rm L} &=& 0, \nonumber \\
D \rho_4^{\rm L} &=& -\left(\mathcal{H}_3+\sqrt{3}\mathcal{H}_8 \right)\rho_5^{\rm L}  -(l_3+\sqrt{3}l_8)^2\rho_4^{\rm L}, \nonumber \\
D \rho_5^{\rm L} &=& \left(\mathcal{H}_3+\sqrt{3}\mathcal{H}_8 \right)\rho_4^{\rm L} -(l_3+\sqrt{3}l_8)^2\rho_5^{\rm L}, \\ 
D \rho_6^{\rm L} &=& -\left(-\mathcal{H}_3+\sqrt{3}\mathcal{H}_8 \right)\rho_7^{\rm L} -(l_3-\sqrt{3}l_8)^2\rho_6^{\rm L}, \nonumber \\ 
D \rho_7^{\rm L} &=& \left(-\mathcal{H}_3+\sqrt{3}\mathcal{H}_8 \right)\rho_6^{\rm L} - (l_3-\sqrt{3}l_8)^2\rho_7^{\rm L}, \nonumber \\
D \rho_8^{\rm L} &=& 0. \nonumber
\end{eqnarray}
Its solution in mass basis acquires the following decaying modes
\begin{eqnarray}
\rho_{{\rm }1}^{\rm L}(x_{21})  & = & \rho_{{\rm }1}(x_{21})\exp\left[{4\int \frac{l_3^2}{2 \mathcal{H}_3} dx_{21}}\right] , \label{eq:lindbladsol1} \\ 
\rho_{{\rm }2}^{\rm L}(x_{21})  & = & \rho_{{\rm }2}(x_{21})\exp\left[{4\int \frac{l_3^2}{2 \mathcal{H}_3} dx_{21}}\right] , \\
\rho_{{\rm }4}^{\rm L}(x_{31})  & = & \rho_{{\rm }4}(x_{31})\exp\left[{\int \frac{(l_3+\sqrt{3}l_8)^2}{\mathcal{H}_3+\sqrt{3}\mathcal{H}_8} dx_{31}}\right] , \\
\rho_{{\rm }5}^{\rm L}(x_{31})  & = & \rho_{{\rm }5}(x_{31})\exp\left[{\int \frac{(l_3+\sqrt{3}l_8)^2}{\mathcal{H}_3+\sqrt{3}\mathcal{H}_8} dx_{31}}\right] , \\
\rho_{{\rm }6}^{\rm L}(x_{32})  & = & \rho_{{\rm }6}(x_{32})\exp\left[{\int \frac{(l_3-\sqrt{3}l_8)^2}{-\mathcal{H}_3+\sqrt{3}\mathcal{H}_8} dx_{32}}\right] , \\
\rho_{{\rm }7}^{\rm L}(x_{32})  & = & \rho_{{\rm }7}(x_{32})\exp\left[{\int \frac{(l_3-\sqrt{3}l_8)^2}{-\mathcal{H}_3+\sqrt{3}\mathcal{H}_8} dx_{32}}\right] , \label{eq:lindbladsol7}
\end{eqnarray}
where the first terms on the right side are identical to the solutions without the Lindblad operator, as in equations (\ref{eq:soltwo1}) -- (\ref{eq:soltwo7}). 
The terms $\rho_{{\rm }0}$, $\rho_{{\rm }3}$ and $\rho_{{\rm }8}$ are constants and equal to their initial condition given by equations (\ref{eq:initerm0}), 
(\ref{eq:initerm3}) and (\ref{eq:initerm8}).

\begin{figure}
\centering
\includegraphics[width=\textwidth]{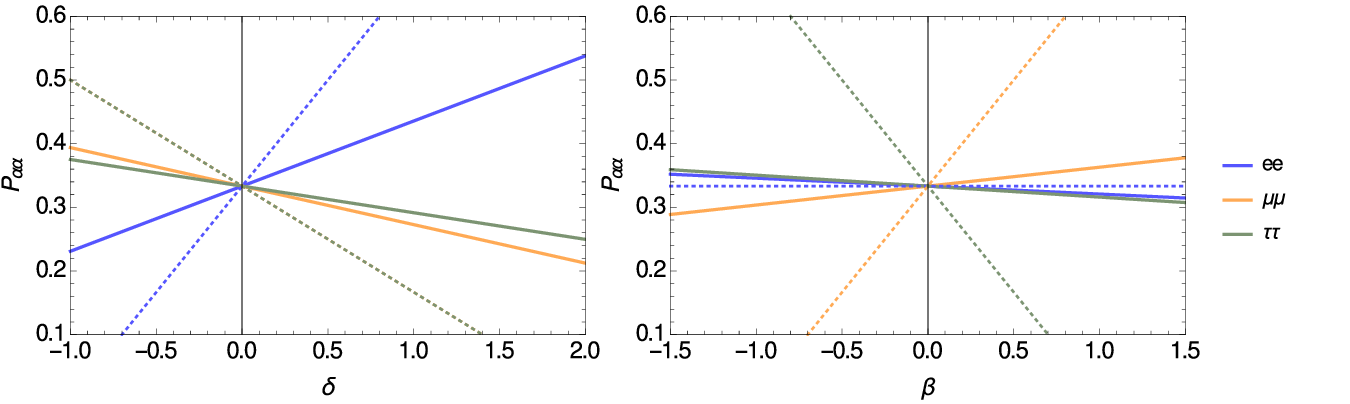}      
\caption{Probabilities to find cosmological neutrinos at different flavour states before (dotted) and after (solid) 
decoherence as a function of the initial distortions $\delta = \delta(q)$ (left, $\beta$ fixed to zero) and $\beta = \beta(q)$ 
(right, $\delta$ fixed to zero). The initial distortion for electron neutrinos is $\delta/3$, for muon neutrinos 
$-\delta/6+\beta/3$, and for tau neutrinos $-\delta/6-\beta/3$. The mixing angles are given by the global fit to 
neutrino oscillation data \cite{Tortola:2012te} for the normal hierarchy and we assume a vanishing CP-violation phase. 
\label{fig3}}
\label{probability}	
\end{figure}

\subsection{Averaging}

The procedure to obtain time-averaged solutions is identical to the two-flavour case. We consider 
that the Lindblad operator acts on the solutions in mass basis, suppressing the time-dependent terms of the density 
matrix (i.e.\ $\rho_{{\rm }1},\rho_{{\rm }2},\rho_{{\rm }4},\rho_{{\rm }5},\rho_{{\rm }6},\rho_{{\rm }7}\rightarrow 0$). 
Then the averaged density matrix in flavour basis is obtained by rotating the remaining time-independent terms 
(i.e.\ $\rho_{{\rm }0},\rho_{{\rm }3},\rho_{{\rm }8}$) to obtain expressions similar to 
(\ref{eq:averagedflavour0}) -- (\ref{eq:averagedflavour3}). 

Applying the Lindblad operator simplifies the solution in the mass basis, but the rotation to the 
most general flavour basis generates solutions that are again too lengthy to include here. 
However, we can show simple solutions using the best-fit values for the mixing angles in normal hierarchy and 
vanishing CP-violation 
\begin{eqnarray}
\bar \rho_{{\rm f}0}  & =  & 1/3,  \label{eq:averflavour0} \\ 
\bar \rho_{{\rm f}1}  & =  & -0.08223 \delta-0.02451 \beta , \label{eq:averflavour1} \\
\bar \rho_{{\rm f}2}  & =  & 0 , \label{eq:averflavour2} \\
\bar \rho_{{\rm f}3}  & =  & 0.08141 \delta-0.02108 \beta , \label{eq:averflavour3} \\
\bar \rho_{{\rm f}4}  & =  & 0.03797 \delta+0.05279 \beta  , \label{eq:averflavour4} \\
\bar \rho_{{\rm f}5}  & =  & 0 , \label{eq:averflavour5} \\
\bar \rho_{{\rm f}6}  & =  & -0.09714 \delta+0.06137 \beta , \label{eq:averflavour6} \\
\bar \rho_{{\rm f}7}  & =  & 0 , \label{eq:averflavour7} \\
\bar \rho_{{\rm f}8}  & =  & 0.03621 \delta+0.01492 \beta . \label{eq:averflavour8}
\end{eqnarray} 
This allows us to obtain the probability $P_{\alpha\alpha}$ to find a neutrino of the CNB
in flavour state $\alpha$,
\begin{eqnarray}
P_{ee} &=& \bar \rho_{{\rm f}0} + \bar \rho_{{\rm f}3} + \frac{1}{\sqrt{3}} \bar \rho_{{\rm f}3} 
= \frac{1}{3} + 0.1023 \delta - 0.0125 \beta, \\
P_{\mu\mu} &=& \bar \rho_{{\rm f}0} - \bar \rho_{{\rm f}3} + \frac{1}{\sqrt{3}} \bar \rho_{{\rm f}3}
= \frac{1}{3} - 0.0605 \delta + 0.0297 \beta, \\
P_{\tau\tau} &=& \bar \rho_{{\rm f}0} - \frac{2}{\sqrt{3}} \bar \rho_{{\rm f}3}
= \frac{1}{3} - 0.0418 \delta - 0.0172 \beta.
\end{eqnarray}
A graphical illustration of this result is provided in figure~\ref{fig3}. It shows that todays CNB 
does not necessarily have a 1:1:1 mix of the three active neutrino flavours. In fact, the expected mix of neutrino 
flavours depends on the initial values of the spectral distortions $\delta(q)$ and $\beta(q)$. In the standard 
(minimal) scenario with vanishing lepton-flavour asymmetries we have $\beta \ll \delta = {\cal O}(10^{-4})$ and 
deviations from flavour equality are tiny.

\subsection{Discussion}

The Lindblad operator introduces two degrees of freedom for the rate of decoherence ($l_3$ and $l_8$) but the 
formalism itself does not indicate when the operator should start to act. 
To better understand the evolution of the three-flavour system, we present the increase of the oscillation phase 
as a function or redshift in figure~\ref{graphentropyn}. We see that the phases evolve in the same way as long as
the neutrinos are effectively relativistic. This may give a first hint that the decoherence of cosmological neutrinos 
should start when the heaviest neutrino becomes non-relativistic, as already discussed for the case of two 
neutrino flavours.

\begin{figure}
\includegraphics[width=0.49\textwidth]{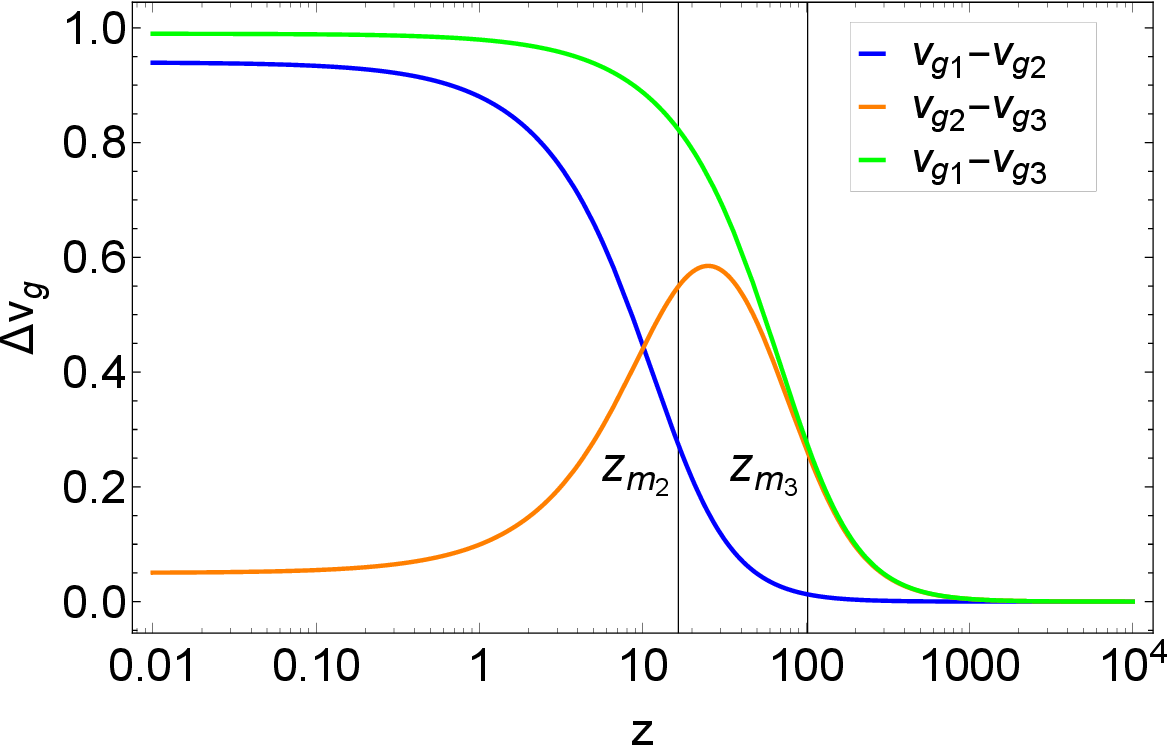}
\includegraphics[width=0.49\textwidth]{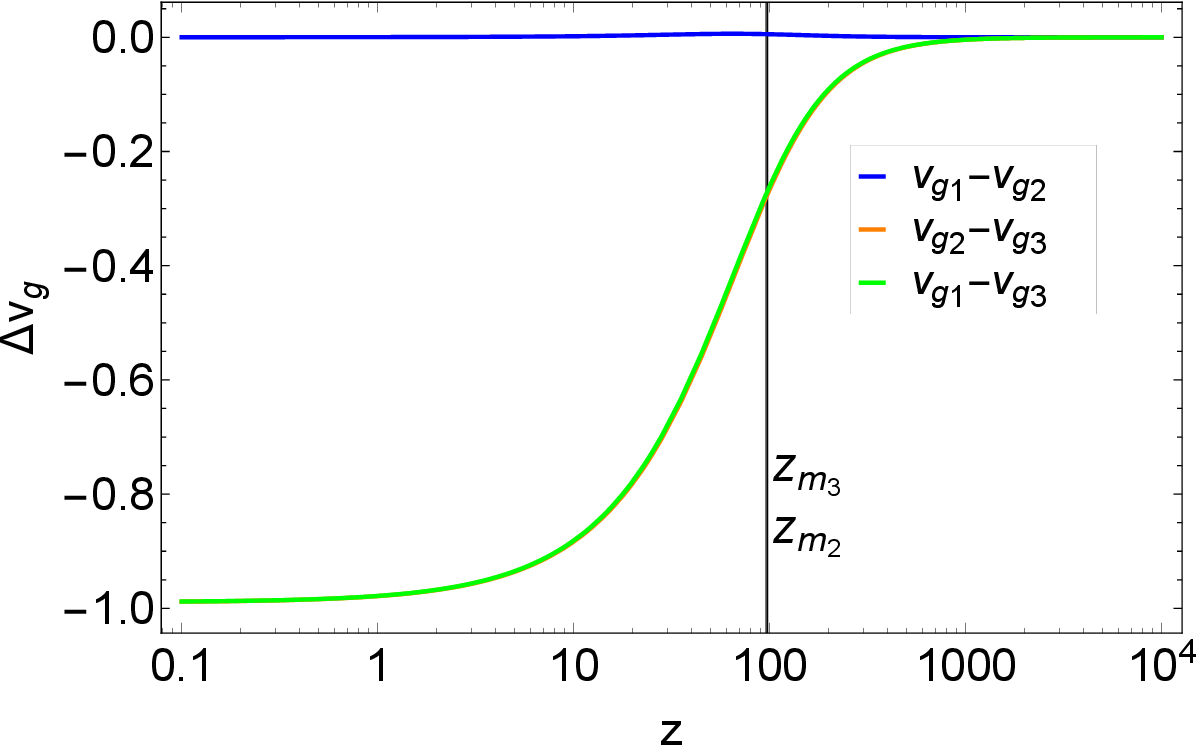}
\caption{Difference of group velocities of neutrinos mass states for normal (left) and inverted (right) hierarchy ($m_{\rm lightest}=0$), shown for neutrinos at the peak of their thermal distribution ($q=3.15T_\nu$). The vertical lines mark the time of transition to the non-relativistic regime for the two massive neutrino states. } 
\label{groupvelocity}	
\end{figure}

We have learned from the discussion of the two-flavour case, that decoherence via a (in general time and momentum 
dependent) Lindblad operator and averaging lead to the same asymptotic density matrix. 
The propagation speed of a wave packet is given by the respective group velocity. In the case of 
three neutrino masses we have three different group velocities, which are functions of redshift,  
\begin{equation} 
v_{{\rm g}i}(z) = \frac{q (1+z)}{\sqrt{m^2_i + q^2 (1+z)^2}}.  
\end{equation} 
In figure~\ref{groupvelocity} we plot the difference of these group velocities for pairs of 
neutrino mass states for $q=3.15T_\nu$. We observe that the group velocities are identical as long as all 
neutrinos are relativistic. They differ when the heaviest neutrino becomes non-relativistic. 
Again, this suggests that coherent neutrino oscillations, as described by the von Neumann equation, 
take place during the relativistic neutrino propagation. Without decoherence that statement would hold until today. 

One could argue that neutrino oscillations can be averaged shortly after neutrino decoupling and thus decoherence takes 
place in the early Universe, however it seems that there is no justification for that, as neutrinos do not scatter 
at $T< 1$ MeV and after the annihilation of positrons and electrons the number of potential scattering partners of neutrinos 
drops by another factor of $10^9$. The fact that the oscillating phase assumes high values does not necessarily 
mean that averaging is due automatically, since in principle one could still recover its exact value and obtain 
the corresponding microscopic state and trace it back to the initial state. A mechanism able to distinguish the mass 
states is still necessary. We suggest here that it is the transition from the relativistic to the 
non-relativistic evolution that induces decoherence of cosmological neutrinos and it is the difference in the inertial mass 
of the different neutrino states that ``couple'' in a different way to space-time. In a perfectly isotropic and 
homogeneous space-time, the different phase and group velocities would lead to a different world lines for the three 
mass states, but instead of evolving as independent ``classical'' states, they would evolve as entangled states. We think that 
situation must be different when we consider that the space-time is homogeneous and isotropic in a statistical sense only. 
We suggest that in such a situation the entanglement of the states will probably be lost due to their (gravitational) 
interaction with the environment.

\section{Entropy evolution}

\subsection{Two-flavour case}

Let us now have a closer look at the family entropy of the system. For density 
matrices with $|\rho_i| \ll 1/2$, we approximate the von Neumann entropy, such that the logarithm can be Taylor expanded,
\begin{eqnarray} 
S &=& - \mathrm{tr} \left[ \left(\frac{1}{2} I + \rho_i \sigma_i\right) \ln \left(\frac{1}{2} I + \rho_j \sigma_j\right) \right] 
      \nonumber \\
       &=& \ln 2 - \frac{1}{2} \mathrm{tr} \left[ \left(I + 2 \rho_i \sigma_i\right) \ln \left(I + 2 \rho_j \sigma_j\right) \right] 
       \nonumber \\
       &=& \ln 2 - 2 \sum_{i=1}^3 \rho_i^2 + \mathcal{O}(4),
\end{eqnarray}  
where we utilise the well-known properties of Pauli matrices, $\mathrm{tr} \sigma_i=0$ and $\mathrm{tr} \sigma_i\sigma_j=2\delta_{ij}$. 
The allowed range of initial conditions gives 
$ \ln 2 - \frac 1 2 \leq S \leq \ln 2$ at the leading order. Presumably, 
higher order corrections change that into $0 \leq S \leq \ln 2$.

We can also obtain an exact expression for the family entropy. In the flavour basis, 
\begin{equation}
S =  \ln2 -\frac{1}{2}(1 - \delta )\ln [1 - \delta ] - \frac{1}{2}(1+\delta)\ln[1+ \delta ] \ .
\end{equation}
For small initial spectral distortions, $|\delta| \ll 1$, we find
\begin{equation}
S(x) = \ln 2 - \frac{\delta^2}{2} + \mathcal{O}(4).
\end{equation}
Thus the family entropy is constant as quantum coherence persists as long as the neutrino oscillations are 
determined by the von Neumann equation.

Decoherence due to whatever reason leads to the increase of entropy.  If we calculate the family entropy from the asymptotic solutions to the Lindblad equation, we find
\begin{eqnarray}
 S = \ln2-\frac{1}{2}(1 - \delta  \cos (2\theta)) \ln [1 - \delta  \cos (2\theta)]- \frac{1}{2}(1+\delta  \cos (2\theta)) \ln [1+\delta  \cos (2\theta)] ,
\end{eqnarray}
while for small distortions
\begin{equation}
S = \ln 2 - \cos^2(2\theta)\frac{\delta^2}{2} + \mathcal{O}\left([\cos(2\theta)\delta]^4\right) \ .
\end{equation}
Therefore, the increase in family entropy caused by decoherence is
\begin{equation}
\Delta S = S_{\rm decoherent} - S_{\rm coherent} = \sin^2(2 \theta) \frac{\delta^2}{2} + \mathcal{O}(\delta^4).
\end{equation} 
It is maximal for maximal mixing, which also results in the maximum entropy state. For maximal mixing and an 
expected spectral distortion $\delta$ of order $10^{-4}$, we find an increase of family entropy of the order of $10^{-8}$.

Let us note that the time average of the entropy without decoherence does not change, as the calculation of 
entropy and  averaging do not commute. Thus the mathematical identity of the time averaged density matrix and the 
asymptotic density matrix after decoherence does not correspond to a physical equivalence of decoherence and time 
averaging. Nevertheless, this mathematical identity can be useful in the calculation and is exploited in this work. 

\subsection{Three-flavour case}

The entropy for three neutrino flavours and small spectral distortions can be Taylor expanded, 
\begin{eqnarray} 
S &=& - \mathrm{tr} \left[ \left(\frac{1}{3} I + \rho_i \lambda_i\right) \ln \left(\frac{1}{3} I + \rho_j \lambda_j\right) \right] 
      \nonumber \\
       &=& \ln 3 - 3\sum_{i=1}^8 \rho_i^2 + \mathcal{O}(4),
\end{eqnarray} 
where we made use of $\mbox{tr} \lambda_i = 0$ and $\mbox{tr} \lambda_i \lambda_j = 2 \delta_{ij}$.
For the initial conditions specified in the previous section, we find the simple result
\begin{equation}
S = \ln 3 - \frac{\delta^2}{4} - \frac{\beta^2}{3} + \mathcal{O}(4).
\end{equation}

We can now calculate the increase of entropy due to decoherence. It is possible to obtain an 
analytical result, dependent on the mixing angles $\theta_{12}$, $\theta_{13}$, $\theta_{23}$
and the initial flavour distortions $\delta$ and $\beta$. For vanishing distortion between muon and tau 
neutrinos ($\beta\rightarrow 0$) there is no dependence on the angle $\theta_{23}$ or in the CP-violation phase 
(complete result in the appendix~\ref{sec:appendixa}). This second degree of freedom ($\beta$) is responsible 
for distinguishing the third-level state, without it the system becomes identical to a two-level system, 
when mixing between second and third state is irrelevant and it is not possible to develop CP-violation.

\begin{figure}
\includegraphics[width=1\textwidth]{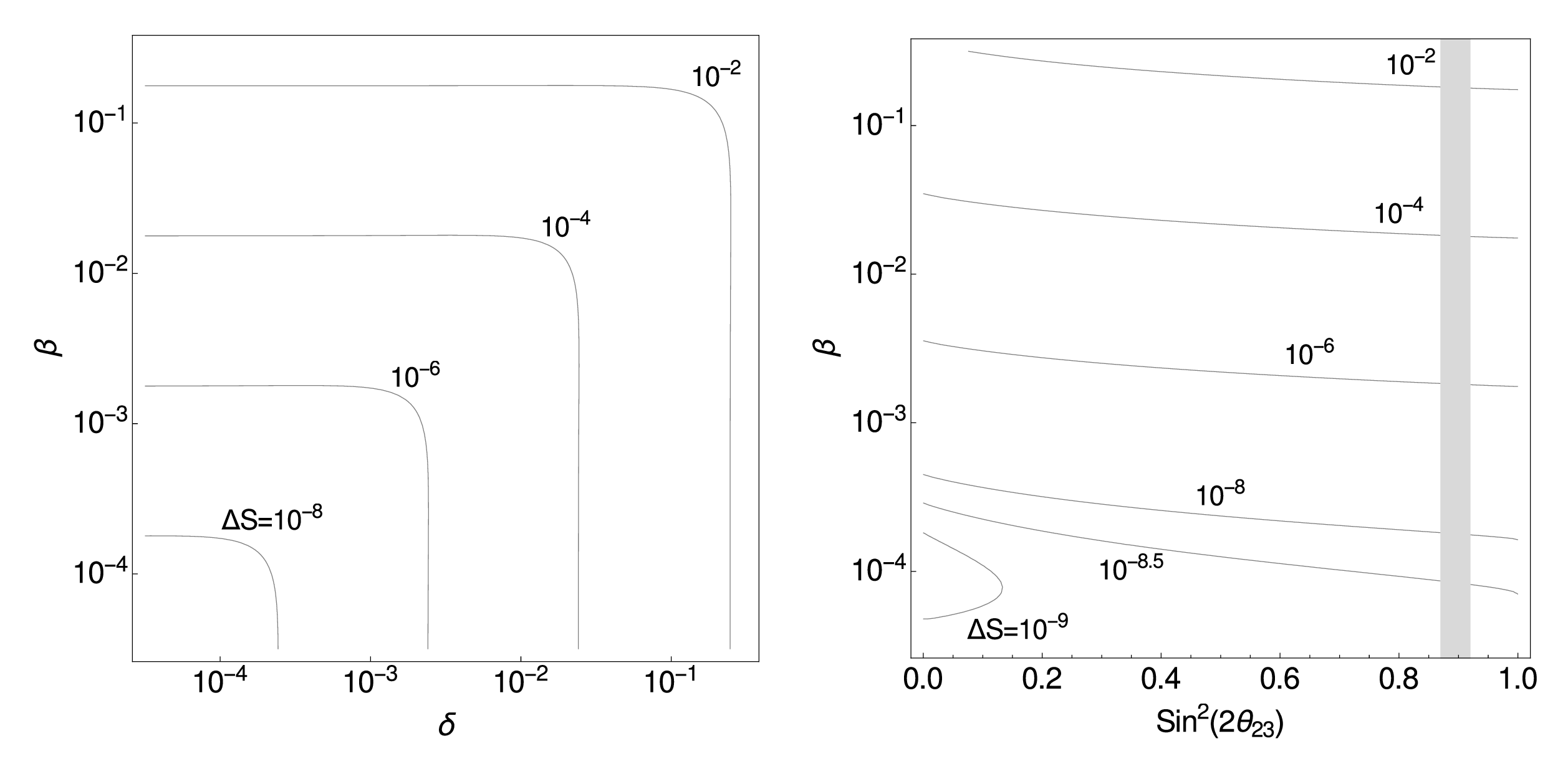}
\caption{Family entropy increase for a system of three mixed neutrino states as functions of the initial spectral distortions 
$\delta$ and $\beta$ (left). The values for the mixing angles are taken from the global fit to neutrino-oscillation data \cite{Tortola:2012te}, assuming a 
normal neutrino hierarchy and a vanishing Dirac CP-violation phase. In the right panel the distortion 
$\delta$ is set to $10^{-4}$ and the mildly constrained mixing angle $\theta_{23}$ is allowed to vary (shaded area is the 
allowed $3\sigma$ region). } \label{graphentropy}	
\end{figure}

In order to calculate the increase in flavour entropy of the neutrino due to decoherence,  we make 
use of the fact that the von Neumann entropy does not depend on the basis in which the density matrix is given; 
$\mbox{tr}[\rho_a \ln(\rho_a)] = \mbox{tr}[\rho_b \ln(\rho_b)]$, where $(a,b)$ are different bases. We choose the 
most suitable basis to calculate the entropy difference of the initial (coherent) states and the final (decoherent) states,
\begin{equation}
\Delta S = S_{\rm decoherent} - S_{\rm coherent} = -\mbox{tr}[\bar\rho\ln(\bar\rho)] +
   \mbox{tr}[\rho_{\rm f}(x_{\rm ini})\ln(\rho_{\rm f}(x_{\rm ini}))] \ .
\end{equation} 
The initial and final entropy is calculated using the flavour and mass basis, respectively, 
for both the density matrix is diagonal in the suitable basis. Here we exploit the mathematical identity of the 
asymptotic solutions for the density matrix from the Lindblad equation and the time averaged solutions of the 
von Neumann equation. 

Although it is possible to obtain an exact solution, we choose to present an approximation for small distortions

\begin{eqnarray}
& \Delta S = \frac{3\delta ^2}{64} \cos^2 \theta_{13} \left[9-\cos (4 \theta_{12})-(7+\cos (4 \theta_{12})) \cos (2 \theta_{13})\right] \nonumber \\
& +\frac{\beta ^2}{192}  \left[64-\cos^2(2 \theta_{23}) \left(2 \cos (4 \theta_{12}) (\cos(2 \theta_{13})-3)^2+12 \cos (2 \theta_{13})+7 \cos (4 \theta_{13})+37\right) \right] \nonumber \\
& -\frac{\beta ^2}{3}  \sin^2(2 \theta_{12}) \sin^2 \theta_{13} \sin^2(2 \theta_{23}) \nonumber \\
& -\frac{\beta ^2}{48}   \sin (4 \theta_{12}) [\sin (3 \theta_{13})-7 \sin \theta_{13}] \sin (4 \theta_{23}) \nonumber \\
& +\frac{\beta  \delta}{16}\cos^2 \theta_{13} \cos (2 \theta_{23}) \left[5-(\cos (4 \theta_{12})+7) \cos (2 \theta_{13})+3 \cos (4 \theta_{12})\right] \nonumber \\
& -\frac{\beta  \delta}{4} \cos^2 \theta_{13} \sin (4 \theta_{12}) \sin \theta_{13} \sin (2 \theta_{23}) + \mathcal{O}(4)\ . \label{eq:changeentropy}
\end{eqnarray}
This is an excellent approximation up to $\delta\sim 0.5$ and, contrary to the exact solution, shows a simple 
dependence on the parameters.

For vanishing primordial lepton-flavour asymmetries, the special case of identical distortion for muon and 
tau neutrinos is theoretically well motivated. Asymmetries between these two flavours are not expected when the 
spectral distortions were produced. 
The combination of the expected initial distortion ($\delta=4.45\times 10^{-4}$, $\beta=0$)
\cite{Mangano:2005cc} with the measured mixing angles
\cite{Tortola:2012te} gives rise to an increase of flavour entropy, $\Delta S = 3.43\times 10^{-8}$. 
In figure~\ref{graphentropy} we present an interesting non-trivial case, when each flavour has a 
different distortion and we show the dependence on the only mildly constrained mixing angle $\theta_{23}$.
We find that the latter affects the cosmological predictions only weakly.

\section{Conclusion}

In this work we have studied cosmological aspects of the decoherence of mixed neutrino states. We have 
described decoherence phenomena via Lindblad operators in the von Neumann equation (\ref{vNnew}). 
 
The evolution of the family composition of the cosmic neutrino background, from the time when neutrinos are 
decoupled until today, has been investigated in section 3. We obtain the expected flavour composition of the CNB 
as a function of arbitrary initial spectral distortions, mixing angles and Dirac CP-violation phase. The net effect
on the current flavour composition is shown in figure~\ref{probability}, where we see that neutrino oscillations and 
the effect of decoherence tend to equilibrate the initial flavour composition. For the measured mixing angles the 
equilibration is not perfect and a small residual flavour imbalance is expected. The remnant (in general 
momentum dependent) spectral distortion of electron neutrinos is expected to be of the order of $10^{-4}$ in the 
minimal scenario (no primeval lepton-flavour asymmetry).

The PTOLEMY experiment \cite{Betts:2013uya} is a proposal to detect cosmological neutrinos by
looking for electron kinetic energies beyond the end point of the tritium $\beta$-decay spectrum 
\cite{PhysRev.128.1457}. According to our result, the fraction of electron neutrinos in the CNB carries information
on the initial spectral distortion after $e^\pm$-annihilation and consequently about the state of the 
universe at that time. One could even speculate about futuristic detectors with such an exquisite sensitivity that even 
CNB intensity anisotropies \cite{Hannestad:2009xu}, similar to the cosmic microwave temperature anisotropies, 
would be detected. The residual imbalance of neutrino flavours in the minimal scenario is one order of
magnitude larger than the expected CNB anisotropies (apart from the dipole). The flavour imbalance would be increased
for a lepton-flavour asymmetric universe.

We obtained exact solutions for the time-dependent Wigner density matrix, valid for any mass and 
momentum in a homogeneous and isotropic universe. The use of Lindblad operators results in 
a similar phenomenology as time-averaging for the asymptotic density 
matrices. We demonstrated that explicitly in section 2 for a two-flavour example.

While it is interesting that Lindblad operators and averaging give rise to the same 
asymptotic results for the density matrix, the time of 
decoherence and the details of its mechanism are not provided by the formalism used. In the three-flavour case we are 
left with two unknown, non-trivial and real functions, $l_3(x)$ and $l_8(x)$. In order to obtain physical intuition, we also studied the behaviour of the oscillation phase and the group velocities and their differences in figures \ref{graphentropyn} 
and \ref{groupvelocity}. We found that the group velocities start to differ significantly once the heaviest neutrino mass state 
becomes non-relativistic. This suggests that the transition to the non-relativistic regime might trigger decoherence in the 
mass basis via the stochastic inhomogeneities of space-time, which are experienced differently by the different mass states. Subsequently neutrinos would propagate in non-degenerate mass states, which means that they are 
in a frozen mix of flavour states. 

The problem of identifying the decoherence time can also be approached from an experimental 
perspective by looking for observables related to the decoherence process. Recently, a similar idea has been 
proposed by Weinberg \cite{Weinberg:2016uml}, where he suggested that decoherence of an atomic three-level 
system in the context of atomic clocks could provide enough information on the time scale of decoherence 
that one could measure or at least constrain the Lindblad coefficients. Likewise, decoherence of cosmological 
neutrinos could produce a traceable observable signal by the increase of CNB entropy that immediately follows 
the decoherence process.

Such an entropy increase of neutrinos due to decoherence of mixed states was pointed out for
supernova neutrinos \cite{roberto} and for cosmological neutrinos \cite{Bernardini:2012uc} previously. 
The latter work assumed a fixed initial distortion proportional to the cross-section
of each neutrino at the time when the oscillations started. 
We went beyond and considered arbitrary spectral distortions and mixing angles. To the best of our knowledge, the 
entropy increase with its dependence on the initial spectral distortions, mixing angles and CP-violation phase 
in the cosmological context is obtained for the first time. This result is valid regardless the time of when the 
decoherence process happens. Therefore, if we could track the CNB entropy as a function of time, we would 
determine the moment of neutrino decoherence observationally.

It remains to figure out how the CNB entropy could actually be measured:  Since the CNB is isotropic, any 
increase in entropy could only manifest itself macroscopically as an induced dissipative pressure (bulk viscosity \cite{Weinberg:1971mx,straumann}). 
Thus an entropy increase due to decoherence would affect the cosmic neutrino equation of state. Any change 
of the equation of state gives rise to a contribution to the integrated Sachs-Wolfe (ISW) effect. It is thus 
interesting to ask if such an effect could be large enough to be observable. 
If decoherence happens when the neutrinos become non-relativistic, i.e., when neutrinos 
develop different group velocities as shown in figure~\ref{groupvelocity}, then 
the decoherence contribution to the ISW effect would show up at $z \gtrsim 100$. However, we expect the effect 
to be tiny, since it is not only suppressed by $\delta^2 \sim 10^{-8}$, but also by the ratio of neutrino density to 
matter  density at $z \gtrsim 100$. Together this gives an effect of order $10^{-10}$ in the temperature anisotropies.
A lepton-flavour asymmetric universe might however give rise to a larger entropy increase 
(see figure \ref{graphentropy}). 

In this work we focused on the minimal scenario and proposed a small residual flavour imbalance 
of the CNB and a tiny increase of neutrino entropy at $z \sim 100$. A more detailed investigation of 
lepton-flavour asymmetric models might reveal useful constraints on the primeval flavour composition of the Universe. \\

{\bf Acknowledgments}
We are thankful to E. Zavanin, I. Oldengott, D. B\"odeker and Y. Wong for discussions on 
neutrino oscillations and fruitful suggestions. We are also grateful to an anonymous referee for 
valuable comments and questions. DB acknowledges the financial support by CAPES-CSF, 
grant 8768-13-7, and by the ``Bielefeld Young Researchers' Fund". HESV thanks CNPq and FAPES. 
We acknowledge the support from the RTG 1620 ``Models of Gravity" funded by DFG.

\bibliographystyle{unsrt}

\appendix
\section{Appendix: Contribution of the CP-violation phase} \label{sec:appendixa}

In this appendix we explore results with non-vanishing CP-violation phase. The results are given
in resemblance with the previous section for the three-flavour case. There is no change in the
dynamics, solely in the mixing matrix. We consider the same initial conditions as in equation
(\ref{eq:initialthreeflavours}) and use the same differential equation (\ref{vNcomponentsthreen}) 
to obtain the results in mass basis. First, we show the solutions for each Gell-Mann block. 
The blocks $\lambda_0$, $\lambda_3$ and $\lambda_8$ are still constant
\begin{eqnarray}
\rho_{{\rm }0}  & =  &  \frac{1}{3} ,  \label{eq:initerm0cp} \\
\rho_{{\rm }3}  & =  & +\frac{\delta}{4} \cos(2 \theta_{12}) \cos^2\theta_{13} \nonumber \\
&& + \frac{\beta}{12} \cos (2 \theta_{12}) (\cos (2 \theta_{13})-3) \cos (2 \theta_{23}) \nonumber \\
&& + \frac{\beta}{3} \sin (2 \theta_{12}) \sin \theta_{13} \sin (2 \theta_{23}) \cos (\delta_{\rm CP}), \label{eq:initerm3cp} \\
\rho_{{\rm }8}  & =  & +\frac{\delta}{8 \sqrt{3}}  \left( 3\cos(2 \theta_{13})-1\right) \nonumber \\
&& +\frac{\beta}{2 \sqrt{3}} \cos^2(\theta_{13}) \cos (2 \theta_{23}) \label{eq:initerm8cp}, 
\end{eqnarray}
while the other blocks are time dependent, but now with dependence on the phase $\delta_{\rm CP}$
\begin{eqnarray}
\rho_{{\rm }1}(x_{21})  & =  &  +\frac{\delta}{4} \sin (2 \theta_{12}) \cos^2 \theta_{13}\cos (x_{21})\nonumber \\
&&-\frac{\beta}{3} \cos (\delta_{\rm CP}) \cos (2 \theta_{12}) \sin \theta_{13} \sin (2 \theta_{23})\cos (x_{21})\nonumber \\
&& +\frac{\beta}{12}\sin (2 \theta_{12}) (\cos (2 \theta_{13})-3) \cos (2\theta_{23})\cos (x_{21})\nonumber \\ 
&& -\frac{\beta}{3} \sin (\delta_{\rm CP}) \sin \theta_{13} \sin (2 \theta_{23}) \sin (x_{21}) , \label{eq:soltwo1a} \\
\rho_{{\rm }2}(x_{21})  & =  &   -\frac{1}{4} \delta  \sin (2 \theta_{12}) \cos^2 \theta_{13}\sin (x_{21})\nonumber \\
&& + \frac{\beta}{3} \cos (\delta_{\rm CP}) \cos (2 \theta_{12}) \sin \theta_{13} \sin (2 \theta_{23})\sin (x_{21})\nonumber\\
&& -\frac{\beta}{12} \sin (2 \theta_{12}) (\cos (2 \theta_{13})-3) \cos (2 \theta_{23})\sin (x_{21})\nonumber \\
&& -\frac{\beta}{3} \sin (\delta_{\rm CP}) \sin \theta_{13} \sin (2 \theta_{23}) \cos (x_{21}), \label{eq:soltwo2a} \\
\rho_{{\rm }4}(x_{31})  & =  &   +\frac{\delta}{4}   \cos \theta_{12} \sin (2 \theta_{13}) \cos (x_{31}-\delta_{\rm CP}) \nonumber \\
&&+\frac{\beta}{6} \cos \theta_{12} \sin (2 \theta_{13}) \cos (2 \theta_{23}) \cos (x_{31}-\delta_{\rm CP}) \nonumber \\
&& -\frac{\beta}{3} \sin \theta_{12} \cos \theta_{13} \sin (2 \theta_{23}) \cos (x_{31}), \label{eq:soltwo4a}  
\end{eqnarray}
\begin{eqnarray}
\rho_{{\rm }5}(x_{31})  & =  &   -\frac{\delta}{4} \cos \theta_{12} \sin (2\theta_{13}) \sin (x_{31}-\delta_{\rm CP})\nonumber \\
&& -\frac{\beta}{6} \cos \theta_{12} \sin (2\theta_{13}) \cos (2 \theta_{23}) \sin (x_{31}-\delta_{\rm CP}) \nonumber \\
&&  +\frac{\beta}{3} \cos \theta_{13}\sin \theta_{12} \sin (2 \theta_{23}) \sin (x_{31}) , \label{eq:soltwo5a} \\
\rho_{{\rm }6}(x_{32})  & =  &   +\frac{\delta}{4} \sin \theta_{12} \sin (2\theta_{13}) \cos (x_{32}-\delta_{\rm CP})\nonumber\\
&&+\frac{\beta}{6} \sin \theta_{12} \sin (2\theta_{13}) \cos (2 \theta_{23}) \cos (x_{32}-\delta_{\rm CP}) \nonumber\\
&& +\frac{\beta}{3} \cos \theta_{13} \cos \theta_{12} \sin (2\theta_{23}) \cos (x_{32}), \label{eq:soltwo6a} \\
\rho_{{\rm }7}(x_{32})  & =  &  -\frac{\delta}{4} \sin \theta_{12} \sin (2\theta_{13}) \sin (x_{32}-\delta_{\rm CP}) \nonumber \\
&& -\frac{\beta}{6} \sin \theta_{12} \sin (2\theta_{13}) \cos (2 \theta_{23}) \sin (x_{32}-\delta_{\rm CP})\nonumber\\
&&  -\frac{\beta}{3} \cos \theta_{13}\cos \theta_{12} \sin (2 \theta_{23}) \sin (x_{32}) \label{eq:soltwo7a} .
\end{eqnarray}
We observe that the phase $\delta_{\rm CP}$ introduces a difference of phase for each solution as 
long as the initial distortion $\beta$ is non-vanishing, otherwise it becomes a global phase that can be 
absorbed in the initial phase.

\begin{figure}
\includegraphics[height=0.26\textheight]{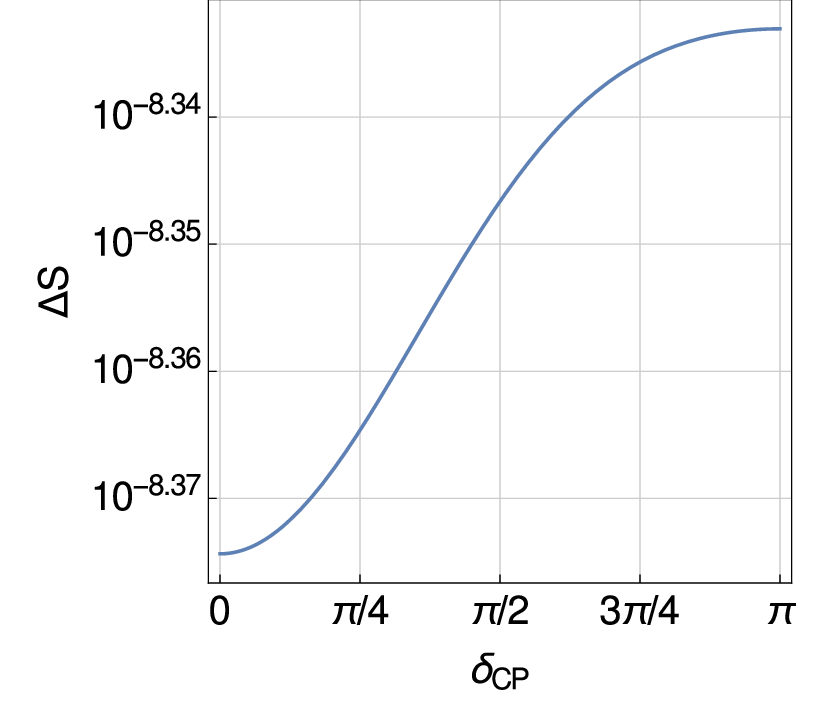} 
\includegraphics[height=0.26\textheight]{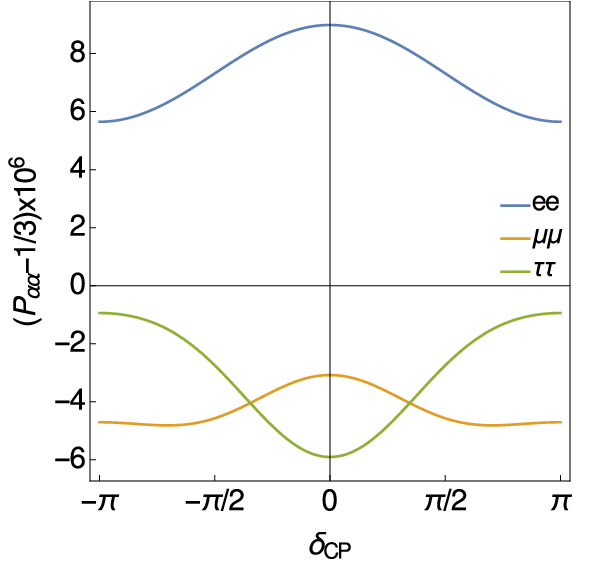}
\caption{Change in entropy (left) and flavour imbalance (right) 
after decoherence as a function of CP-violation phase. Initial spectral distortions $\delta$ and $\beta$ are both 
fixed to the value $10^{-4}$. The values for the mixing angles are given by the global fit of neutrino oscillation 
data \cite{Tortola:2012te} for normal hierarchy.}
\label{graphentropyCP}	
\end{figure}

For non-vanishing CP-violation phase, the difference between probabilities for neutrinos and anti-neutrinos in non-trivial. In 
the case of the transition from electron to muon neutrinos it is given by
\begin{eqnarray}
 P_{e\mu}-P_{\bar e  \bar \mu} &=& \sin (2 \theta_{12}) \sin (2 \theta_{23}) \cos ^2(\theta_{13}) \sin (\theta_{13})\sin (\delta_{\rm CP}) \nonumber \\
 \times && \left[\sin (x_{21})\exp\left({4\int \frac{l_3^2}{2 \mathcal{H}_3} dx_{21}}\right) \right. \nonumber \\
 && -\sin (x_{31}) \exp\left({\int \frac{(l_3+\sqrt{3}l_8)^2}{\mathcal{H}_3+\sqrt{3}\mathcal{H}_8} dx_{31}}\right) \nonumber \\
 && \left. +\sin (x_{32})\exp\left({\int \frac{(l_3-\sqrt{3}l_8)^2}{-\mathcal{H}_3+\sqrt{3}\mathcal{H}_8} dx_{32}}\right) \right] ,
\end{eqnarray}
which is consistent with the literature \cite{PhysRevLett.45.2084}. In order to stress the role of coherence in this 
particular result, we use the Gell-Mann coefficients for the solutions contained in equations (\ref{eq:lindbladsol1}) -- (\ref{eq:lindbladsol7}).
We note that differences between neutrinos and anti-neutrinos for any transition probability vanish in the limit of 
completed decoherence.

We obtain the limit after decoherence is completed of equations (\ref{eq:initerm0cp}) -- (\ref{eq:soltwo7a}) and
then rotate to the flavour basis. For simplicity we replace the mixing angles for the best-fit in the global 
analysis \cite{Tortola:2012te} for normal neutrino hierarchy
\begin{eqnarray}
\bar \rho_{{\rm f}0}  & =  & 1/3,  \\
\bar \rho_{{\rm f}1}  & =  & -\delta[0.05032 + 0.03192 \cos (\delta_{\rm CP})] \nonumber \\
&& - \beta[0.00900 + 0.01447 \cos (\delta_{\rm CP}) + 0.00104 \cos (2 \delta_{\rm CP})]  , \\
\bar \rho_{{\rm f}2}  & =  & -0.03192 \delta  \sin (\delta_{\rm CP}) + \beta[0.01276 \sin (\delta_{\rm CP})-0.00104 \sin (2 \delta_{\rm CP})]  , \\
\bar \rho_{{\rm f}3}  & =  & \delta  [0.08767 -0.00626 \cos (\delta_{\rm CP})] \nonumber \\
&& - \beta[0.02992-0.01053 \cos (\delta_{\rm CP})+0.00169 \cos (2 \delta_{\rm CP})] , \\
\bar \rho_{{\rm f}4}  & =  & \delta  [0.06333 -0.02536 \cos (\delta_{\rm CP})] \nonumber \\
&&+\beta  [0.00920 + 0.04442 \cos (\delta_{\rm CP})-0.00082 \cos (2 \delta_{\rm CP})]  , \\
\bar \rho_{{\rm f}5}  & =  & -0.02536 \delta  \sin (\delta_{\rm CP}) \nonumber \\
&& - \beta  [0.01014 \sin (\delta_{\rm CP})+0.00082 \sin (2 \delta_{\rm CP})] ,\\
\bar \rho_{{\rm f}6}  & =  & +\delta  [0.09423+0.00290 \cos (\delta_{\rm CP})] \nonumber \\
&& +\beta  [0.05407 + 0.00808 \cos (\delta_{\rm CP})-0.00079 \cos (2 \delta_{\rm CP})] \\
\bar \rho_{{\rm f}7}  & =  & 0.00551 \beta  \sin (\delta_{\rm CP})+0.01285 \delta  \sin (\delta_{\rm CP}) ,\\
\bar \rho_{{\rm f}8}  & =  & \delta  [0.02539 + 0.01084 \cos (\delta_{\rm CP})] \nonumber \\
&& + \beta  [0.00133 + 0.01065 \cos (\delta_{\rm CP})+0.00293 \cos (2 \delta_{\rm CP})] ,
\end{eqnarray} 
which are consistent with the solution for vanishing CP-violation phase present in equations (\ref{eq:averflavour0}) -- (\ref{eq:averflavour8}). 

Using the averaged solution in mass basis for simplicity, we can calculate the change in the von Neumann entropy 
using equation (\ref{eqentropy})
\begin{eqnarray}
& \Delta S = \frac{3\delta ^2}{64} \cos^2 \theta_{13} \left[9-\cos (4 \theta_{12})-(7+\cos (4 \theta_{12})) \cos (2 \theta_{13})\right] \nonumber \\
& +\frac{\beta ^2}{192}  \left[64-\cos^2(2 \theta_{23}) \left(2 \cos (4 \theta_{12}) (\cos(2 \theta_{13})-3)^2+12 \cos (2 \theta_{13})+7 \cos (4 \theta_{13})+37\right) \right] \nonumber \\
& -\frac{\beta ^2}{3}  \cos^2(\delta_{\rm CP}) \sin^2(2 \theta_{12}) \sin^2 \theta_{13} \sin^2(2 \theta_{23}) \nonumber \\
& -\frac{\beta ^2}{48}  \cos (\delta_{\rm CP}) \sin (4 \theta_{12}) [\sin (3 \theta_{13})-7 \sin \theta_{13}] \sin (4 \theta_{23}) \nonumber \\
& +\frac{\beta  \delta}{16}\cos^2 \theta_{13} \cos (2 \theta_{23}) \left[5-(\cos (4 \theta_{12})+7) \cos (2 \theta_{13})+3 \cos (4 \theta_{12})\right] \nonumber \\
& -\frac{\beta  \delta}{4}\cos (\delta_{\rm CP}) \cos^2 \theta_{13} \sin (4 \theta_{12}) \sin \theta_{13} \sin (2 \theta_{23}) + \mathcal{O}(4)\ .
\end{eqnarray}
The change in entropy is also consistent with the case of a vanishing CP-violation phase in 
equation (\ref{eq:changeentropy}), and identical for vanishing initial difference between muon and 
tau ($\beta=0$). In figure~\ref{graphentropyCP} we show the probability to find a cosmological neutrino in a specified 
flavour state  in the averaged limit as well as the expected change in the entropy, both as functions of the 
CP-violation phase.

\end{document}